\documentclass[superscriptaddress,twocolumn,reprint]{revtex4-1}

\usepackage[version=3]{mhchem} % Formula subscripts using \ce{}
\usepackage{array}		% need for special property functions
\usepackage{amsmath}    % need for subequations
\usepackage{supertabular}
\usepackage{graphicx}   % need for figures
\usepackage{verbatim}   % useful for program listings
\usepackage{color}      % use if color is used in text
\usepackage{subfigure}  % use for side-by-side figures
\usepackage{epspdfconversion} % convert EPS to PDF on the fly
\usepackage{hyperref}   % use for hypertext links, including those to external documents and URLs

\begin{document}

\title{Phase Transformation Dynamics in Porous Battery Electrodes}
\author{Todd R. Ferguson}
\affiliation{Department of Chemical Engineering, Massachusetts Institute of Technology}
\author{Martin Z. Bazant}
\affiliation{Department of Chemical Engineering, Massachusetts Institute of Technology}
\affiliation{Department of Mathematics, Massachusetts Institute of Technology}
\date{\today}

\begin{abstract}
Porous electrodes composed of multiphase active materials are widely used in Li-ion batteries, but their dynamics are poorly understood.  Two-phase models are largely empirical, and no models exist for three or more phases.   Using a modified porous electrode theory based on non-equilibrium thermodynamics, we show that experimental phase behavior can be accurately predicted from free energy models, without artificially placing phase boundaries or fitting the open circuit voltage.     First, we simulate lithium intercalation in porous iron phosphate, a popular two-phase cathode,  and show that the zero-current voltage gap, sloping voltage plateau and under-estimated exchange currents all result from size-dependent nucleation and mosaic instability.   Next, we simulate porous graphite, the standard anode with three stable phases,  and reproduce experimentally observed fronts of color-changing  phase transformations.  These results provide a framework for physics-based design and control for electrochemical systems with complex thermodynamics.   
\end{abstract}

\maketitle

%\section{Introduction}

\section{ Introduction }
In order to develop safer, longer-lasting, and more energy-dense batteries, it is crucial to understand their thermodynamic behavior out of equilibrium.   Many battery materials exhibit multiple phases with varying composition, voltage, and temperature~\cite{handbookbatterymaterials,whittingham2004,bruce2008,dunn2011}, driven by electrochemical reactions~\cite{bazant2013}.   In Li-ion batteries, complex phase transformations are triggered by lithium intercalation reactions.  The standard anode material, graphite, passes through three or more phases ~\cite{ohzuku1993} with observable color changes~\cite{harris2010}, while popular cathode materials, such as olivine phosphates~\cite{padhi1997,tarascon2001,morgan2004,tang2010} and transition metal oxides~\cite{hwang2003,park2007}, exhibit two-phase separation across single particles~\cite{chen2006,ramana2009} and porous electrodes~\cite{chueh2013}.  Conversion reactions can also lead to complex phase behavior in lithium-sulfur batteries~\cite{cheon2003,dominko2011,kumaresan2008} and lithium-oxygen batteries~\cite{lu2012,gallant2012a}.  

Phase transformations pose a major challenge for mathematical models used to design, characterize, and control Li-ion batteries~\cite{ramadesigan2012}.  Systems-level models neglect phase transformations altogether and rely on empirical constructs, such as current-dependent particle sizes~\cite{safari2009,delacourt2011}. Classical porous electrode theory (PET) captures the microscopic physics of diffusion and reactions~\cite{srinivasan2004,newman1975,newman_book}, but does not consistently describe multiphase thermodynamics~\cite{lai2010,lai2011a,bazant2013,ferguson2012}. In all cases, the voltage plateau for two coexisting phases is fitted to a unique (single-phase) function of the state of charge. For two-phase  materials, such as Li$_X$FePO$_4$ (LFP, $X\approx0$, $1$)~\cite{padhi1997}, spherical ``shrinking core" ~\cite{srinivasan2004,dargaville2010} or planar~\cite{cwang2007,kasavajjula2008,zhu2010} phase boundaries are imposed in the active particles, but no such models are available for materials, such as graphite, with three or more phases. 

Recent work on LFP has shown that the voltage plateau is an emergent property of the active particles  ~\cite{tang2010,vanderven2009,bai2011,cogswell2012,lai2011b,dargaville2013CHR,zeng2013MRS,zeng2014} and the porous electrode ~\cite{dreyer2010,dreyer2011,ferguson2012,bai2013,levi2013} for any material with multiple stable phases at different compositions.  
In single particles, phase separation occurs within the miscibility gap (range of unstable homogeneous compositions), which depends on temperature and particle size~\cite{tang2010,cogswell2012}, overpotential~\cite{meethong2007, tang2009, tang2010,cogswell2013}, and current density~\cite{bai2011,cogswell2012}.  In porous electrodes, a collection of particles entering the spinodal gap (linearly unstable compositions) can also undergo discrete transformations~\cite{dreyer2010,dreyer2011,chueh2013,burch_thesis}, but the roles of  nucleation~\cite{cogswell2013,bai2013}, ion transport~\cite{ferguson2012,dargaville2013porous}, and heterogeneity~\cite{bai2013,chueh2013} are just beginning to be understood.

In this work, we present a predictive theory of phase transformations in Li-ion batteries. The mathematical framework, combining classical porous electrode theory~\cite{newman_book} with non-equilibrium  thermodynamics~\cite{bazant2013},  is described in a companion paper ~\cite{ferguson2012}.  This new approach effectively homogenizes microstructural simulations~\cite{garcia2005,garcia2007}, even with phase separation~\cite{orvananos2013}, at a tiny fraction of the computational cost. Here, we introduce simple free energy models  for LFP (two phases with coherent nucleation) and graphite (three phases, neglecting nucleation) and use our modified porous electrode theory (MPET) to predict representative experimental data~\cite{dreyer2010,harris2010}.

%  Although our conclusions also follow from the full model, we present simulations and analysis of the ``pseudocapacitor" limit of homogenenous intercalation in each particle due to fast diffusion or suppressed phase separation compared to experimental time scales. 

\begin{figure*}[t]
\centering	
	\includegraphics[width=6.5in,keepaspectratio]{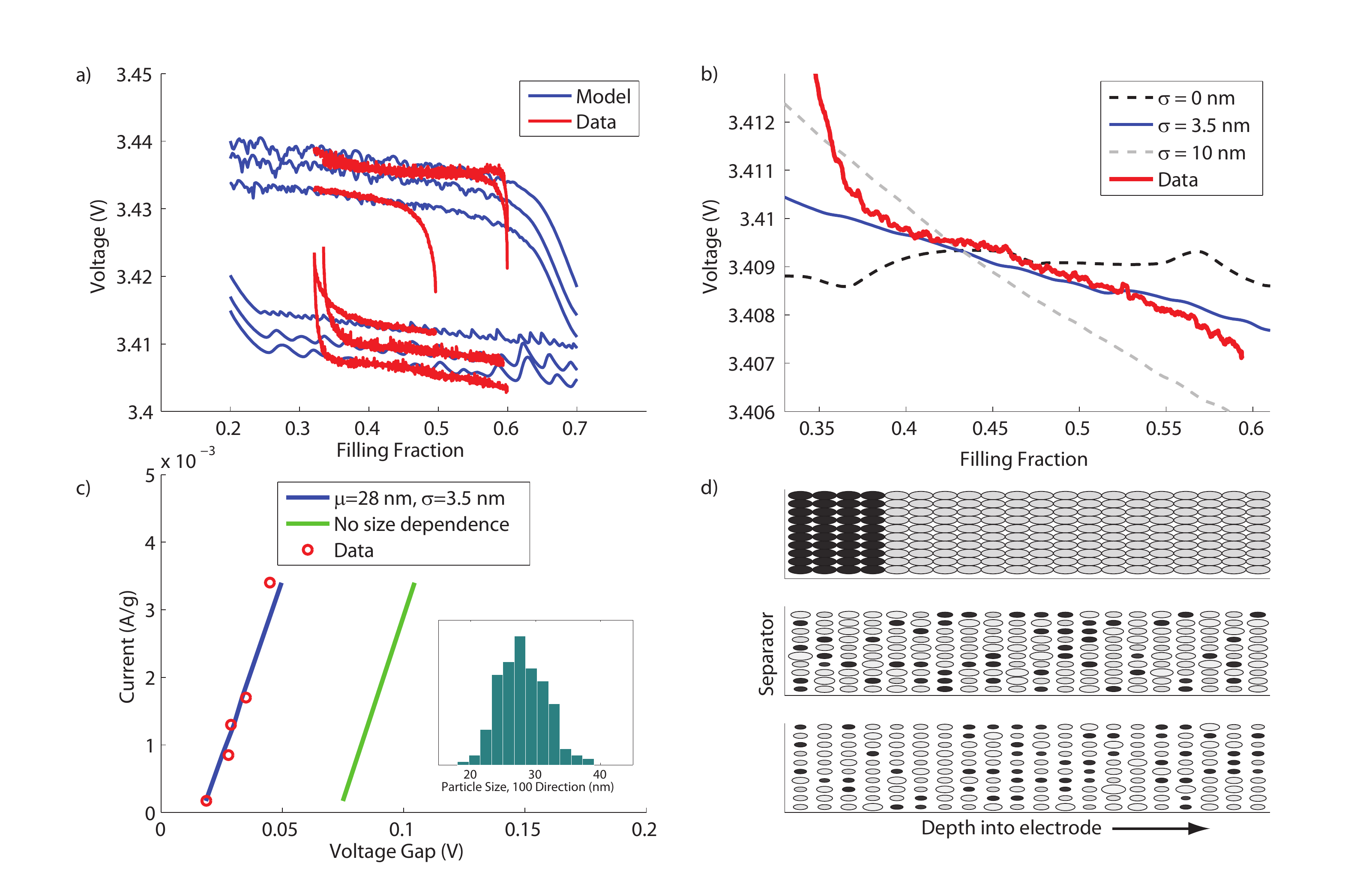}
\caption{  Simulations with modified porous electrode theory for  two-phase Li$_x$FePO$_4$, compared to the experimental data of Dreyer et al. ~\cite{dreyer2010}.  (a) Charge-discharge cycles at C/1000, C/200, and C/131. (b) Predicted effect of the particle-size standard deviation $\sigma$ (in a log-normal distribution) on the discharge curve at C/200 for the same mean size $\mu=28$nm. The simulation results are smoothed; the raw data appears in Fig.~\ref{fig:nosmooth}. (c) Charge-discharge voltage gap, with additional C/10 and C/50 experimental data. Theoretical curves are shown for a single particle size ($\sigma=0$) and for the fitted size distribution in the inset. (d) Simulated lithium profiles across the porous cathode at 50\% filling for the cases in (b), showing the transition from a sharp reaction front at $\sigma=0$ to homogeneous filling in order of increasing particle size with increasing $\sigma$. Supporting Information contains the simulation movies.
}
\label{figure1}
\end{figure*}

%\section{Reaction-Limited Dynamics}

\section{  Iron Phosphate:  Two Phases }   
The understanding of binary phase separation in single particles is rapidly advancing~\cite{bazant2013,tang2010}.  Phase-field models of LFP have been developed for isotropic spherical particles~\cite{tang2009,tang2010,kao2010,zeng2013MRS,zeng2014}, which may be relevant for large particles or agglomerates, while new modes of intercalation have been identified~\cite{bazant2013,dargaville2013CHR} for anisotropic single-crystal nanoparticles~\cite{morgan2004,islam2005}.    At low currents, intercalation waves sweep across the active (010) surface~\cite{singh2008,delmas2008,bai2011,tang2011} and relax to striped patterns along \{100\}~\cite{chen2006} or \{101\}~\cite{ramana2009} planes due to coherency strain~\cite{cogswell2012}.  Once the local current density exceeds the exchange rate, phase separation is suppressed~\cite{bai2011,cogswell2012}.  A ``single-phase transition path" has also been suggested based on the large barrier for bulk nucleation~\cite{malik2011}, but there is new evidence (strengthened below) that two-phase nucleation does occur, via the instability of surface layers~\cite{cogswell2013}. The nucleation barrier is size-dependent and vanishes below a critical particle size around 22nm~\cite{cogswell2013}, consistent with experiments~\cite{ichitsubo2013}.

Much less is known about binary porous electrodes. The seminal work of Dreyer et al. demonstrated a zero-current gap between voltage plateaus for charging and discharging~\cite{dreyer2010}, which they attributed to discrete spinodal decompositions  among bi-stable homogeneous particles, analogous to filing a balloon array~\cite{dreyer2011}.  Independently, Burch~\cite{burch_thesis} observed the ``mosaic instability" in simulations of a collection of phase-separating particles described by the Cahn-Hilliard-reaction model~\cite{burch2009,bazant2013,zeng2013MRS,zeng2014}, which exchange ions through an electrolyte reservoir at constant total current.  (See Ch. 9 of Ref. \cite{burch_thesis}.) Recently, Bai and Tian (BT) interpreted current transients in terms of population dynamics for nucleation and growth in a set of discrete particles~\cite{bai2013} in place of the Kolmogorov-Johnnson-Mehl-Avrami statistical theory for a continuous electrode~\cite{oyama2012}. The BT theory has already led to accurate measurements of mechanical deformation~\cite{levi2013} and charge-transfer kinetics~\cite{bai2014}, and the mosaic picture is directly supported by the first porous-electrode imaging experiments of Chueh et al.~\cite{chueh2013}. The data reveal mostly single-phase particles ($x \approx 0,1$) and some two-phase particles with planar (not core-shell) phase boundaries, consistent with theory~\cite{singh2008,bai2011,cogswell2012,cogswell2013}.  Macroscopic phase gradients are also observed~\cite{chueh2013}, but have never been modeled. 

Here, we use MPET~\cite{ferguson2012,bazant2013,dargaville2013porous} to  re-interpret the data of Dreyer et al.~\cite{dreyer2010} at very low C-rates, C/$n$, discharging the capacity in $n=$ 50-1000 hours.  The porous cathode is partitioned into finite volumes, each containing a representative LFP particle of random size (log-normally distributed) and a realistic shape (C3)~\cite{chen2006,smith2012,cogswell2013}. The separator and lithium anode (at constant potential) are also modeled.     The homogeneous free energy 
\begin{equation}
\overline{g}(x) = k_BT\left[x\ln x +(1-x)\ln(1-x)\right] +\Omega x(1-x)   \label{eq:regsol}
\end{equation}
and diffusional chemical potential
\begin{equation}
\overline{\mu} = \frac{d \overline{g}}{d x} = k_BT\ln\left(\frac{x}{1-x}\right)+\Omega (1-2x)
\end{equation}
(per site) describe a regular solution of particles and vacancies with mean interaction energy $\Omega$, previously fitted to Li-solubility data versus temperature and particle size~\cite{cogswell2012}. 

Based on porous electrode imaging~\cite{chueh2013} and short diffusion times ($\sim$ ms) in nanoparticles~\cite{malik2010}, we assume fast single-particle transformations compared to the C-rate and do not resolve the internal dynamics of each particle.  Instead, we replace each particle with an effective homogeneous solid solution (the ``pseudo capacitor approximation"~\cite{ferguson2012}), whose electrochemical response is tuned to the results of a realistic phase-field model~\cite{cogswell2013}. In particular, the regular solution parameter $\Omega$ is varied with particle size to match the spinodal to the coherent nucleation voltage for discharging (Li insertion)~\cite{cogswell2013}, predicted from {\it ab initio} surface energies~\cite{wang2007} and elastic constants~\cite{maxisch2006}. (See Appendix A.) As a first approximation, the same size-dependent nucleation voltage is also used for charging.  The local Nernst equilibrium voltage (relative to the lithium metal anode) is defined by the chemical potential,
$V_{eq} = V^\Theta - \overline{\mu}/e$,
relative to a formal reference voltage $V^\Theta$ at $x=1/2$ in the middle of the  plateau~\cite{bazant2013,ferguson2012,bai2011}. The local overpotential $\eta=V - V_{eq}$ determines the reaction rate via generalized Butler-Volmer kinetics  from the local interfacial voltage $\Delta\phi$ ~\cite{bazant2013}.    Ion transport is governed by a standard Nernst-Planck electrolyte model~\cite{ferguson2012,newman_book}.  The simulations follow the experimental charge/discharge protocol. 

Our MPET has only three fitting parameters: the mean ($\mu=28$nm) and standard deviation ($\sigma=3.5$nm) in a log-normal distribution of the cycled particle size (defined by the short axis of the C3 shape in the [100] direction, which is roughly half of the long axis length) and a series resistance ($R_s=3.9 \Omega\cdot$g).  Unlike traditional models, the voltage plateau is not fitted, but predicted, and phase transformations occur spontaneously.   Material properties of LFP were previously determined by {\it ab initio} calculations and experiments~\cite{cogswell2012,cogswell2013}, except for the exchange current (see below).  All porous electrode properties are known  or  estimated (electrolyte diffusivity, active material loading, porosity, thickness, tortuosity), but do not significantly affect the results at low rates.

With so few adjustable parameters, the agreement between the theory and experiment in Fig. \ref{figure1}(a) is remarkable.  The discharge plateaus are reproduced with millivolt accuracy, including a slight tilt that had escaped notice. The charging plateaus are also well described, considering that a separate model was not developed for charging nucleation. (The charging data also strangely overlaps for C/200 and C/131, separate from C/1000.)  

An unexpected finding is the effect of particle-size variance on the battery voltage. For a single particle size, the voltage remains constant during phase transformation at zero current. In a realistic heterogeneous composite, the voltage plateau tilts as particles fill in order of increasing nucleation overpotential~\cite{cogswell2013}, from smallest to largest.  Consequently, the zero-current voltage profile can be used to infer the particle size distribution, either by simulations (Fig.  \ref{figure1}) or from a simple analytical formula derived in Appendix C.  
As shown in Fig. \ref{figure1}(b), the theory predicts that only the smallest of the active particles were cycled in these experiments, having  $\mu=28$ nm and $\sigma=3.5$ nm.  The reported range of 50 to 1000 nm~\cite{dreyer2010} may have reflected agglomerates or imaging resolution, since the smaller values also lead to accurate predictions of the voltage gap between charge and discharge versus current, as shown in Fig. \ref{figure1}(c).  Besides collapsing literature data for current transients~\cite{cogswell2013}, the theory of size-dependent coherent nucleation here explains why the observed zero-current gap of 20 mV is much smaller than the bulk spinodal gap of 74 mV inferred from thermodynamic data~\cite{cogswell2012}.  

Another unexpected finding is the effect of particle-size variance on the spatial profile of the phase transformation, shown in Fig. \ref{figure1}(d).  With identical particles ($\sigma=0$), the mosaic instability is localized in a thin reaction front propagating from the separator, whose width is proportional to the current, and voltage fluctuations are enhanced by blocks of transforming particles~\cite{ferguson2012,bazant2013}. With non-identical particles, a small standard deviation (as small as $\sigma = 1$ nm, not shown) suffices to spread the phase transformation over the entire electrode and smooth the voltage profile. The particles fill in spatially extended groups according to size, starting with the smallest, with a gentle gradient in concentration away from the separator.  This theoretical picture is consistent with the imaging experiments of Chueh et al.~\cite{chueh2013}.

These findings can drastically alter the interpretation of electrochemical data for multiphase porous electrodes.  Classical methods~\cite{bard_book}, such as intermittent titration or impedance spectroscopy~\cite{barsoukov2005}, are based on reaction-diffusion models that predict spatially uniform reactions at low current. For solid-solution materials, such as silicon nanowires, this assumption is justified and leads to very accurate impedance  analysis~\cite{song2013}, but for phase-separating materials, mosaic instability implies that  low currents may only probe small fractions of the total active area and particle-size distribution. The reaction rate can be grossly under-estimated if the full internal surface area is assumed to be active.  Indeed, classical PET yields very small exchange current densities (3$\times 10^{-6}$  ~\cite{srinivasan2004} and 5.4$\times 10^{-5}$ \cite{dargaville2010} A/m$^2$), while our MPET uses 7$\times 10^{-3}$ A/m$^2$, which seems more reasonable for a high-rate cathode.  Much larger values $>10$ A/m$^2$ have also been reported ~\cite{cwang2007,kasavajjula2008} that are $>10^7$ times the PET values, so it is clearly important to account for the mosaic instability using MPET.  Since the particle-size variance promotes uniform electrode intercalation,  our results show that the simple BT population-dynamics model~\cite{bai2013} can also be a good approximation at low rates and/or early times (without electrolyte depletion), e.g. to replace the Cottrell equation~\cite{bard_book} in chronoamperometry to more accurately determine the exchange current~\cite{bai2014}.

It is important to stress that our successful fitting of MPET is for low-rate experiments, probing the slow dynamics of electrode phase transformations.   At high rates, the original MPET formulation with Butler-Volmer kinetics and no statistical hetereogeneities~\cite{ferguson2012} can have difficulty fitting experimental data~\cite{dargaville2013porous}. The framework of MPET is very general, however, and it is likely that some more effects must be added to describe the full range of operating conditions.  This work shows that a distribution of particle sizes should be considered, while other studies show the importance of electron transport limitation~\cite{thomas2008} and higher-resistance Marcus-Hush-Chidsey kinetics of electron transfer from the carbon coating of LFP nanoparticles~\cite{bai2014}.

%%%%%%%%%%%%%%%%%%%%%%%%%%%%%%%%%%%%%%%%%%%%%%%%%%%%%%%%%%%%%%%%%%%%%%%%%%%%%%%
%\section{Diffusion Limited Dynamics}

\section{ Graphite:  Three Phases }
Traditional battery models cannot describe active materials with three or more phases, and yet many insertion compounds  have this property.   Empirical approaches, such as ``shrinking annuli"~\cite{hess2013} that generalize the two-phase 	``shrinking core", are difficult to justify theoretically and require inconsistently fitting the open circuit voltage ``staircase" (set of plateaus for multiple phase transformations) as a unique function of the mean concentration.  On the other hand, our MPET has no such limitations and simply requires fitting a homogeneous free energy model with three or more local minima to the phase diagram; the voltage staircase is then an emergent, rate-dependent property of the material. 

\begin{figure*}[t]
\centering	
	\includegraphics[width=6.5in,keepaspectratio]{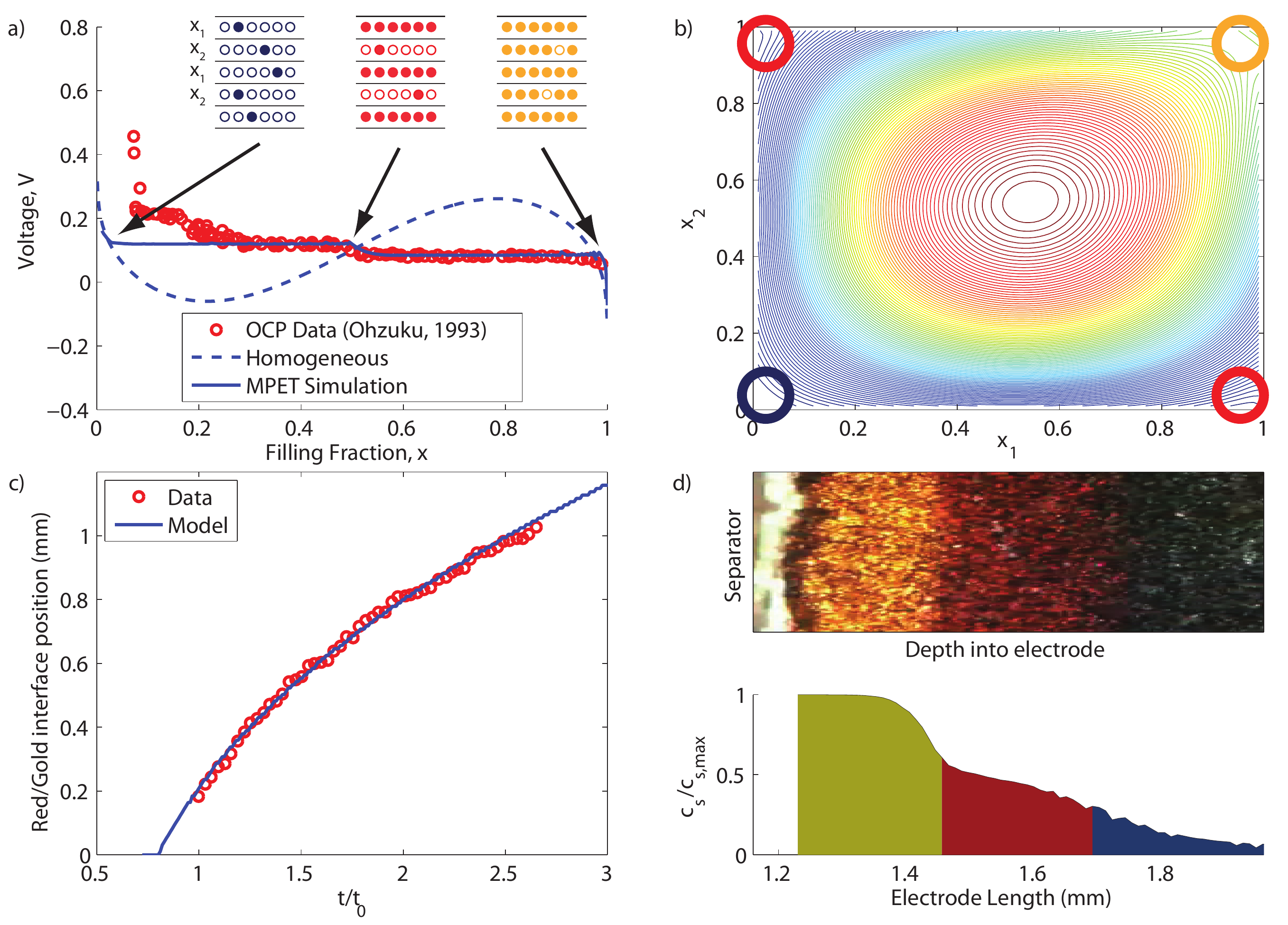}
\caption{ Graphite MPET simulations vs. experiment.  (a) Open circuit voltage ``staircase" for lithium insertion in graphite (vs. Li metal)~\cite{ohzuku1993}, reproduced by the MPET model (solid curve), whose homogeneous free energy is the dashed curve. (b) Contour plot of the free energy model versus the filling fractions of two adjacent, periodically repeating layers  with repulsive interactions across layers.  The model has local minima near (0,0) for  the empty phase (blue), (1,1) for stage I (gold) and (1,0) or (0,1) for stage II (red). These minima are connected by common tangent constructions for two-phase coexistence, which force the system from empty to stage II to stage I during filling. (c) Simulated red/gold interface versus time (solid curve) in the experiments of Harris et al. ~\cite{harris2010}, compared to the experimental data.  (d) Experimental image (above) and simulated color profile (below) at the same moment in time, capturing the blue/red and red/gold interface positions and noise. A movie of the simulation synchronized with the experiment as in (d) is in the supporting information. }
\label{figure2}
\end{figure*}

To illustrate this new approach, we apply MPET to the important case of graphite, a complex intercalation compound~\cite{dresselhaus1981} that has at least five distinct phases during lithium insertion, Li$_x$C$_6$, as $x$ varies from 0 to 1 ~\cite{ohzuku1993}.  
The open circuit voltage (Fig. \ref{figure2}(a)) exhibits two wide plateaus from $x \approx 1/4$ to $x\approx 1/2$ (stage II) to $x\approx 1$ (stage I),  corresponding to the stable states with every third, second, and single layer between graphene planes filled with lithium, respectively.  Higher stages with $0 < x < 1/4$ involve correlated interlayer domains~\cite{krishnan2013} that  are harder to discern experimentally and less important for intercalation dynamics.  As  first approximation, therefore, we consider only the three phases at $x\approx 0$, $x\approx 1/2$ and $x\approx 1$, using a simple free-energy model proposed by Bazant~\cite{10.626}. 

In order to capture two voltage plateaus between these states, the model introduces {\it two order parameters}, $(x_1,x_2)$, that represent the lithium fractions in a pair of periodically repeated layers, where $x = (x_1+x_2)/2$. The inset of Fig. \ref{figure2}(a) illustrates this physical picture of the three phases at the endpoints of the voltage plateaus. The free energy per site in the layer pair, 
\begin{equation}
g(x_1,x_2) = \overline{g}(x_1) + \overline{g}(x_2) + g_{int}(x_1,x_2),
\end{equation}
has local minima near $(1,1)$ for stage I ($x\approx 1$) and $(0,0)$ for the low density phase ($x\approx 0$), as well as equivalent minima near $(1,0)$ and $(0,1)$ for stage II ($x\approx 1/2$), as shown in Fig. \ref{figure2}(b).  Each layer $i=1,2$ has a double-welled homogeneous free energy, $\overline{g}(x_i)$, with minima near $x_i=0,1$ favoring full or empty states. Stage II with every other layer full is stabilized by a repulsive interaction, $g_{int}(x_1,x_2)$, between adjacent layers. 

Building on the LFP model above, each layer is treated as a regular solution,
\begin{equation}
\overline{g}(x_i) = k_BT\left[x_i \ln x_i +(1-x_i)\ln(1-x_i)\right] +\Omega_a x_i(1-x_i)
\end{equation}
where $\Omega_a>0$ controls the width of the each voltage plateau. The interaction energy has a similar polynomial form
\begin{equation}
g_{int}(x_1,x_2) = \Omega_b x_1 x_2+\Omega_c x_1(1-x_1)x_2(1-x_2)
\end{equation}
where $\Omega_b> 0$ is a repulsion energy between two particles at the same site in adjacent layers, which sets the different between the two voltage plateaus.  The second term with $\Omega_c>0$ (not in the original model~\cite{10.626}) is an additional repulsion between adjacent particle-vacancy dipoles, which penalizes partially filled layers and further stabilizes stage II with $x_1\neq x_2$.
Intercalation reactions are allowed to proceed into each layer independently, as if it were a separate reactant. Each layer thus has its own local overpotential, $\eta_i = V - V_{eq,i}$, and Nernst voltage, 
$V_{eq,i} = V^\Theta - \mu_i/e$, 
defined by its diffusional chemical potential,
\begin{equation}
\mu_i = \frac{\partial g}{\partial x_i} = \overline{\mu}_i +\Omega_bx_j+\Omega_cx_j(1-x_j)(1-2x_i)  \ \ \ (i\neq j)
\label{chempotlayeri}
\end{equation}
where 
\begin{equation}
\overline{\mu}_i = \frac{d \overline{g}}{d x_i} = k_BT\ln\left(\frac{x_i}{1-x_i}\right)+\Omega_a (1-2x_i).
\end{equation}
The free-energy model has four parameters, $\Omega_a$, $\Omega_b$, $\Omega_c$ and $V^\Theta$, that are fitted to reproduce the open circuit voltage at a given temperature (Fig. \ref{figure2}(a)). 

Graphite is an electrochromic material that changes color upon lithiation~\cite{dahn1991,nailmova1995,zheng1995}, making it possible to visualize its phase transformations. The low density state is black and switches to blue from a small value of $x< 0.05$ up to $x\approx 1/4$. Stage II is red and extends just past $x\approx 1/2$, while stage I is gold and covers the widest voltage plateau up to $x=1$. These colors are indicated by circles around the corresponding minima of the model free energy in Fig.\ref{figure2}(b).  Slow lithium insertion follows the path of lowest free energy of the convex hull, including common tangent planes between the local minima that represent two-phase co-existence. Lithiation starts near the blue circle (homogeneous empty state) and progresses towards one of the minima inside the red circles (replacing the empty state with stage II) then towards the minimum inside the gold circle (replacing stage II with stage I).  The contour plot also shows the tilt of the free energy along the (1,1) direction, which is controlled by $\Omega_b$ and leads to the different voltage plateau values.

The experiments of Harris et al.~\cite{harris2010} visualizing coloration during lithium intercalation in an ``unrolled" porous graphite electrode provide a unique opportunity to test our MPET model for a material with more than two phases, for the first time. Beautiful experimental movies (posted at http://lithiumbatteryresearch.com) show the nucleation of a sequence of thin reaction fronts propagating diffusively from the separator into the porous electrode during lithium insertion, switching the color of the graphite particles from black to blue to red to gold in discrete, stochastic transformations (Fig.\ref{figure2})(d).  The battery is discharged at a constant potential of 2 mV, very close to short circuit, which leads to a large initial flux of lithium into the particles. 

Since the characteristic time for diffusion across this long electrode  ($\approx 1$ mm)  is on the order of hours, the electrolyte quickly becomes depleted of salt. The propagation of the reaction front is thus limited by the lithium diffusion that feeds it (from both sides), leading to the same square root of time scaling of the front position and diffusion layer width~\cite{harris2010}, as in diffusion-limited corrosion of porous electrodes~\cite{leger1999,bazant2000}. Relative to electrolyte diffusion, reactions and solid diffusion are very fast, so the pseudo-capacitor approximation can be safely made in our MPET~\cite{ferguson2012}, as in the case of LFP nanoparticles above.  Ohmic losses for electrons can also be neglected since the porous graphite layer is very thin and sits on the current collector.  

The MPET simulations also include the separator and lithium metal anode and apply the same constant potential (2mV) as in the experiments. In order to compare with the experimental images, the solid lithium concentration is converted to three colors, consistent with the free energy model and experimental observations:  blue $0 \leq x < 0.3$, red $0.3 \leq x < 0.6$, and gold $0.6 \leq x < 1$.  See Appendix B  for further details.

With essentially no adjustable parameters, the agreement between theory and experiment in Fig.~\ref{figure2}(c)-(d) is remarkable.  Once the free energy model is determined by the open circuit voltage, the only fitting parameter is the effective diffusivity $D_{eff}$ of the electrolyte (since both reactions and solid diffusion are fast), but its value must remain close to theoretical estimates. Assuming the Bruggeman relation, $D_{eff} = \epsilon_p^{3/2} D$, with an ambipolar diffusivity $D=4.6\times10^{-10}$ m$^2$/s (consistent with experimental values for ethylene carbonate / dimethyl carbonate  solution~\cite{stewart2008}),  an excellent fit of the position of the red/gold interface versus time (Fig.~\ref{figure2}(c)) is obtained with a reasonable porosity, $\epsilon_p=0.4$.   This value is somewhat large for a bulk porous electrode, but the experiments involve roughly a monolayer of graphite particles between flat plates, which would have higher porosity.

Fitting the diffusive motion of a single interface may not seem so surprising  (and can also be done by an {\it ad hoc} diffusion equation for coloration~\cite{harris2010}), but our MPET accurately predicts the nucleation and propagation of two stochastic reaction fronts, the red/gold and blue/red interfaces, over the entire recorded time without fitting any additional parameters. (See the supporting movie and Fig.~\ref{figure2}(d).) Moreover, a close examination of the reaction fronts shows that the mosaic instability, i.e. stochastic filling of discrete particles~\cite{dreyer2010,burch_thesis,bai2013,ferguson2012}, is quite similar in both movies, from experiment and simulation.

\section{ Conclusion }

In this paper, we show that MPET~\cite{ferguson2012} based on electrochemical non-equilibrium thermodynamics~\cite{bazant2013} is able to accurately simulate two fundamental experiments with multiphase porous electrodes~\cite{dreyer2011,harris2010}  that traditional porous electrode theories could not describe.  The advantage of MPET is that it couples the thermodynamics of the active material  to electrochemical transport and reaction kinetics.  Complex dynamical phenomena, such as nucleation, phase growth, mosaic instability, and voltage hysteresis, are then predicted by the model, rather than artificially imposed on the system.  

The fundamental input to MPET is a free energy model for the active material, inferred from the phase diagram and open circuit voltage. The cell voltage is predicted as an emergent property of the porous electrode, rather than fitted to experimental data as an effective property of the active material, as if it remained homogeneous and never phase separated. By properly accounting for non-equilibrium thermodynamics, MPET provides a promising framework for battery modeling to optimize the cell design, predict and control performance, monitor the internal state, and improve safety under diverse operating conditions.

\section*{Acknowledgements}
This work was supported by the National Science Foundation
under Contract DMS-0948071 and by the Samsung-MIT Alliance.  The  authors thank  M. Gaber\u{s}\u{c}ek and J. Mo\u{s}kon for sharing experimental data~\cite{dreyer2010}, D. A. Cogswell for sharing simulation results~\cite{cogswell2013}, and Raymond Smith for comments and suggestions.

\section*{Additional Information}  Movies of the simulations in Figures \ref{figure1} and \ref{figure2} are available online.

\bibliography{elec41}

%merlin.mbs apsrev4-1.bst 2010-07-25 4.21a (PWD, AO, DPC) hacked
%Control: key (0)
%Control: author (8) initials jnrlst
%Control: editor formatted (1) identically to author
%Control: production of article title (-1) disabled
%Control: page (0) single
%Control: year (1) truncated
%Control: production of eprint (0) enabled
\begin{thebibliography}{82}%
\makeatletter
\providecommand \@ifxundefined [1]{%
 \@ifx{#1\undefined}
}%
\providecommand \@ifnum [1]{%
 \ifnum #1\expandafter \@firstoftwo
 \else \expandafter \@secondoftwo
 \fi
}%
\providecommand \@ifx [1]{%
 \ifx #1\expandafter \@firstoftwo
 \else \expandafter \@secondoftwo
 \fi
}%
\providecommand \natexlab [1]{#1}%
\providecommand \enquote  [1]{``#1''}%
\providecommand \bibnamefont  [1]{#1}%
\providecommand \bibfnamefont [1]{#1}%
\providecommand \citenamefont [1]{#1}%
\providecommand \href@noop [0]{\@secondoftwo}%
\providecommand \href [0]{\begingroup \@sanitize@url \@href}%
\providecommand \@href[1]{\@@startlink{#1}\@@href}%
\providecommand \@@href[1]{\endgroup#1\@@endlink}%
\providecommand \@sanitize@url [0]{\catcode `\\12\catcode `\$12\catcode
  `\&12\catcode `\#12\catcode `\^12\catcode `\_12\catcode `\%12\relax}%
\providecommand \@@startlink[1]{}%
\providecommand \@@endlink[0]{}%
\providecommand \url  [0]{\begingroup\@sanitize@url \@url }%
\providecommand \@url [1]{\endgroup\@href {#1}{\urlprefix }}%
\providecommand \urlprefix  [0]{URL }%
\providecommand \Eprint [0]{\href }%
\providecommand \doibase [0]{http://dx.doi.org/}%
\providecommand \selectlanguage [0]{\@gobble}%
\providecommand \bibinfo  [0]{\@secondoftwo}%
\providecommand \bibfield  [0]{\@secondoftwo}%
\providecommand \translation [1]{[#1]}%
\providecommand \BibitemOpen [0]{}%
\providecommand \bibitemStop [0]{}%
\providecommand \bibitemNoStop [0]{.\EOS\space}%
\providecommand \EOS [0]{\spacefactor3000\relax}%
\providecommand \BibitemShut  [1]{\csname bibitem#1\endcsname}%
\let\auto@bib@innerbib\@empty
%</preamble>
\bibitem [{han(2011)}]{handbookbatterymaterials}%
  \BibitemOpen
  \href@noop {} {\emph {\bibinfo {title} {Handbook of Battery Materials}}},\
  \bibinfo {edition} {2nd}\ ed.\ (\bibinfo  {publisher} {Wiley},\ \bibinfo
  {year} {2011})\BibitemShut {NoStop}%
\bibitem [{\citenamefont {Whittingham}(2004)}]{whittingham2004}%
  \BibitemOpen
  \bibfield  {author} {\bibinfo {author} {\bibfnamefont {M.~S.}\ \bibnamefont
  {Whittingham}},\ }\href@noop {} {\bibfield  {journal} {\bibinfo  {journal}
  {Chem. Rev.}\ }\textbf {\bibinfo {volume} {104}},\ \bibinfo {pages} {4271}
  (\bibinfo {year} {2004})}\BibitemShut {NoStop}%
\bibitem [{\citenamefont {Bruce}\ \emph {et~al.}(2008)\citenamefont {Bruce},
  \citenamefont {Scrosati},\ and\ \citenamefont {Tarascon}}]{bruce2008}%
  \BibitemOpen
  \bibfield  {author} {\bibinfo {author} {\bibfnamefont {P.~G.}\ \bibnamefont
  {Bruce}}, \bibinfo {author} {\bibfnamefont {B.}~\bibnamefont {Scrosati}}, \
  and\ \bibinfo {author} {\bibfnamefont {J.-M.}\ \bibnamefont {Tarascon}},\
  }\href@noop {} {\bibfield  {journal} {\bibinfo  {journal} {Angewandte
  Chemie}\ }\textbf {\bibinfo {volume} {47}},\ \bibinfo {pages} {2930 }
  (\bibinfo {year} {2008})}\BibitemShut {NoStop}%
\bibitem [{\citenamefont {Dunn}\ \emph {et~al.}(2011)\citenamefont {Dunn},
  \citenamefont {Kamath},\ and\ \citenamefont {Tarascon}}]{dunn2011}%
  \BibitemOpen
  \bibfield  {author} {\bibinfo {author} {\bibfnamefont {B.}~\bibnamefont
  {Dunn}}, \bibinfo {author} {\bibfnamefont {H.}~\bibnamefont {Kamath}}, \ and\
  \bibinfo {author} {\bibfnamefont {J.-M.}\ \bibnamefont {Tarascon}},\
  }\href@noop {} {\bibfield  {journal} {\bibinfo  {journal} {Science}\ }\textbf
  {\bibinfo {volume} {334}},\ \bibinfo {pages} {928} (\bibinfo {year}
  {2011})}\BibitemShut {NoStop}%
\bibitem [{\citenamefont {Bazant}(2013)}]{bazant2013}%
  \BibitemOpen
  \bibfield  {author} {\bibinfo {author} {\bibfnamefont {M.~Z.}\ \bibnamefont
  {Bazant}},\ }\href@noop {} {\bibfield  {journal} {\bibinfo  {journal}
  {Accounts of Chemical Research}\ }\textbf {\bibinfo {volume} {46}},\ \bibinfo
  {pages} {1144} (\bibinfo {year} {2013})}\BibitemShut {NoStop}%
\bibitem [{\citenamefont {Ohzuku}\ \emph {et~al.}(1993)\citenamefont {Ohzuku},
  \citenamefont {Iwakoshi},\ and\ \citenamefont {Sawai}}]{ohzuku1993}%
  \BibitemOpen
  \bibfield  {author} {\bibinfo {author} {\bibfnamefont {T.}~\bibnamefont
  {Ohzuku}}, \bibinfo {author} {\bibfnamefont {Y.}~\bibnamefont {Iwakoshi}}, \
  and\ \bibinfo {author} {\bibfnamefont {K.}~\bibnamefont {Sawai}},\
  }\href@noop {} {\bibfield  {journal} {\bibinfo  {journal} {J. Electrochem.
  Soc.}\ }\textbf {\bibinfo {volume} {140}},\ \bibinfo {pages} {2490} (\bibinfo
  {year} {1993})}\BibitemShut {NoStop}%
\bibitem [{\citenamefont {Harris}\ \emph {et~al.}(2010)\citenamefont {Harris},
  \citenamefont {Timmons}, \citenamefont {Baker},\ and\ \citenamefont
  {Monroe}}]{harris2010}%
  \BibitemOpen
  \bibfield  {author} {\bibinfo {author} {\bibfnamefont {S.~J.}\ \bibnamefont
  {Harris}}, \bibinfo {author} {\bibfnamefont {A.}~\bibnamefont {Timmons}},
  \bibinfo {author} {\bibfnamefont {D.~R.}\ \bibnamefont {Baker}}, \ and\
  \bibinfo {author} {\bibfnamefont {C.}~\bibnamefont {Monroe}},\ }\href@noop {}
  {\bibfield  {journal} {\bibinfo  {journal} {Chemical Physics Letters}\
  }\textbf {\bibinfo {volume} {485}},\ \bibinfo {pages} {265} (\bibinfo {year}
  {2010})}\BibitemShut {NoStop}%
\bibitem [{\citenamefont {Padhi}\ \emph {et~al.}(1997)\citenamefont {Padhi},
  \citenamefont {Nanjundaswamy},\ and\ \citenamefont {Goodenough}}]{padhi1997}%
  \BibitemOpen
  \bibfield  {author} {\bibinfo {author} {\bibfnamefont {A.}~\bibnamefont
  {Padhi}}, \bibinfo {author} {\bibfnamefont {K.}~\bibnamefont
  {Nanjundaswamy}}, \ and\ \bibinfo {author} {\bibfnamefont {J.}~\bibnamefont
  {Goodenough}},\ }\href@noop {} {\bibfield  {journal} {\bibinfo  {journal}
  {Journal of the Electrochemical Society}\ }\textbf {\bibinfo {volume}
  {144}},\ \bibinfo {pages} {1188} (\bibinfo {year} {1997})}\BibitemShut
  {NoStop}%
\bibitem [{\citenamefont {Tarascon}\ and\ \citenamefont
  {Armand}(2001)}]{tarascon2001}%
  \BibitemOpen
  \bibfield  {author} {\bibinfo {author} {\bibfnamefont {J.}~\bibnamefont
  {Tarascon}}\ and\ \bibinfo {author} {\bibfnamefont {M.}~\bibnamefont
  {Armand}},\ }\href@noop {} {\bibfield  {journal} {\bibinfo  {journal}
  {Nature}\ }\textbf {\bibinfo {volume} {414}},\ \bibinfo {pages} {359}
  (\bibinfo {year} {2001})}\BibitemShut {NoStop}%
\bibitem [{\citenamefont {Morgan}\ \emph {et~al.}(2004)\citenamefont {Morgan},
  \citenamefont {der Ven},\ and\ \citenamefont {Ceder}}]{morgan2004}%
  \BibitemOpen
  \bibfield  {author} {\bibinfo {author} {\bibfnamefont {D.}~\bibnamefont
  {Morgan}}, \bibinfo {author} {\bibfnamefont {A.~V.}\ \bibnamefont {der Ven}},
  \ and\ \bibinfo {author} {\bibfnamefont {G.}~\bibnamefont {Ceder}},\
  }\href@noop {} {\bibfield  {journal} {\bibinfo  {journal} {Electrochemical
  and Solid State Letters}\ }\textbf {\bibinfo {volume} {7}},\ \bibinfo {pages}
  {A30} (\bibinfo {year} {2004})}\BibitemShut {NoStop}%
\bibitem [{\citenamefont {Tang}\ \emph {et~al.}(2010)\citenamefont {Tang},
  \citenamefont {Carter},\ and\ \citenamefont {Chiang}}]{tang2010}%
  \BibitemOpen
  \bibfield  {author} {\bibinfo {author} {\bibfnamefont {M.}~\bibnamefont
  {Tang}}, \bibinfo {author} {\bibfnamefont {W.~C.}\ \bibnamefont {Carter}}, \
  and\ \bibinfo {author} {\bibfnamefont {Y.-M.}\ \bibnamefont {Chiang}},\
  }\href@noop {} {\bibfield  {journal} {\bibinfo  {journal} {Annual Review of
  Materials Research}\ }\textbf {\bibinfo {volume} {40}},\ \bibinfo {pages}
  {501} (\bibinfo {year} {2010})}\BibitemShut {NoStop}%
\bibitem [{\citenamefont {Hwang}\ \emph {et~al.}(2003)\citenamefont {Hwang},
  \citenamefont {Tsai}, \citenamefont {Carlier},\ and\ \citenamefont
  {Ceder}}]{hwang2003}%
  \BibitemOpen
  \bibfield  {author} {\bibinfo {author} {\bibfnamefont {B.~J.}\ \bibnamefont
  {Hwang}}, \bibinfo {author} {\bibfnamefont {Y.~W.}\ \bibnamefont {Tsai}},
  \bibinfo {author} {\bibfnamefont {D.}~\bibnamefont {Carlier}}, \ and\
  \bibinfo {author} {\bibfnamefont {G.}~\bibnamefont {Ceder}},\ }\href@noop {}
  {\bibfield  {journal} {\bibinfo  {journal} {Chemistry of Materials}\ }\textbf
  {\bibinfo {volume} {15}},\ \bibinfo {pages} {3676} (\bibinfo {year}
  {2003})}\BibitemShut {NoStop}%
\bibitem [{\citenamefont {Park}\ \emph {et~al.}(2007)\citenamefont {Park},
  \citenamefont {Kang}, \citenamefont {Johnson}, \citenamefont {Amine},\ and\
  \citenamefont {Thackeray}}]{park2007}%
  \BibitemOpen
  \bibfield  {author} {\bibinfo {author} {\bibfnamefont {S.}~\bibnamefont
  {Park}}, \bibinfo {author} {\bibfnamefont {S.}~\bibnamefont {Kang}}, \bibinfo
  {author} {\bibfnamefont {C.}~\bibnamefont {Johnson}}, \bibinfo {author}
  {\bibfnamefont {K.}~\bibnamefont {Amine}}, \ and\ \bibinfo {author}
  {\bibfnamefont {M.}~\bibnamefont {Thackeray}},\ }\href@noop {} {\bibfield
  {journal} {\bibinfo  {journal} {Electrochemistry Communications}\ }\textbf
  {\bibinfo {volume} {9}},\ \bibinfo {pages} {262} (\bibinfo {year}
  {2007})}\BibitemShut {NoStop}%
\bibitem [{\citenamefont {Chen}\ \emph {et~al.}(2006)\citenamefont {Chen},
  \citenamefont {Song},\ and\ \citenamefont {Richardson}}]{chen2006}%
  \BibitemOpen
  \bibfield  {author} {\bibinfo {author} {\bibfnamefont {G.}~\bibnamefont
  {Chen}}, \bibinfo {author} {\bibfnamefont {X.}~\bibnamefont {Song}}, \ and\
  \bibinfo {author} {\bibfnamefont {T.}~\bibnamefont {Richardson}},\
  }\href@noop {} {\bibfield  {journal} {\bibinfo  {journal} {Electrochemical
  and Solid State Letters}\ }\textbf {\bibinfo {volume} {9}},\ \bibinfo {pages}
  {A295} (\bibinfo {year} {2006})}\BibitemShut {NoStop}%
\bibitem [{\citenamefont {Ramana}\ \emph {et~al.}(2009)\citenamefont {Ramana},
  \citenamefont {Mauger}, \citenamefont {Gendron}, \citenamefont {Julien},\
  and\ \citenamefont {Zaghib}}]{ramana2009}%
  \BibitemOpen
  \bibfield  {author} {\bibinfo {author} {\bibfnamefont {C.~V.}\ \bibnamefont
  {Ramana}}, \bibinfo {author} {\bibfnamefont {A.}~\bibnamefont {Mauger}},
  \bibinfo {author} {\bibfnamefont {F.}~\bibnamefont {Gendron}}, \bibinfo
  {author} {\bibfnamefont {C.~M.}\ \bibnamefont {Julien}}, \ and\ \bibinfo
  {author} {\bibfnamefont {K.}~\bibnamefont {Zaghib}},\ }\href@noop {}
  {\bibfield  {journal} {\bibinfo  {journal} {J. Power Sources}\ }\textbf
  {\bibinfo {volume} {187}},\ \bibinfo {pages} {555} (\bibinfo {year}
  {2009})}\BibitemShut {NoStop}%
\bibitem [{\citenamefont {Chueh}\ \emph {et~al.}(2013)\citenamefont {Chueh},
  \citenamefont {Gabaly}, \citenamefont {Sugar}, \citenamefont {Bartelt},
  \citenamefont {McDaniel}, \citenamefont {Fenton}, \citenamefont {Zavadil},
  \citenamefont {Tyliszczak}, \citenamefont {Lai},\ and\ \citenamefont
  {McCarty}}]{chueh2013}%
  \BibitemOpen
  \bibfield  {author} {\bibinfo {author} {\bibfnamefont {W.~C.}\ \bibnamefont
  {Chueh}}, \bibinfo {author} {\bibfnamefont {F.~E.}\ \bibnamefont {Gabaly}},
  \bibinfo {author} {\bibfnamefont {J.~D.}\ \bibnamefont {Sugar}}, \bibinfo
  {author} {\bibfnamefont {N.~C.}\ \bibnamefont {Bartelt}}, \bibinfo {author}
  {\bibfnamefont {A.~H.}\ \bibnamefont {McDaniel}}, \bibinfo {author}
  {\bibfnamefont {K.~R.}\ \bibnamefont {Fenton}}, \bibinfo {author}
  {\bibfnamefont {K.~R.}\ \bibnamefont {Zavadil}}, \bibinfo {author}
  {\bibfnamefont {T.}~\bibnamefont {Tyliszczak}}, \bibinfo {author}
  {\bibfnamefont {W.}~\bibnamefont {Lai}}, \ and\ \bibinfo {author}
  {\bibfnamefont {K.~F.}\ \bibnamefont {McCarty}},\ }\href {\doibase
  10.1021/nl3031899} {\bibfield  {journal} {\bibinfo  {journal} {Nano Letters}\
  }\textbf {\bibinfo {volume} {13}},\ \bibinfo {pages} {866 } (\bibinfo {year}
  {2013})}\BibitemShut {NoStop}%
\bibitem [{\citenamefont {Cheona}\ \emph {et~al.}(2003)\citenamefont {Cheona},
  \citenamefont {Koa}, \citenamefont {Choa}, \citenamefont {Kimb},
  \citenamefont {Chinc},\ and\ \citenamefont {Kimd}}]{cheon2003}%
  \BibitemOpen
  \bibfield  {author} {\bibinfo {author} {\bibfnamefont {S.-E.}\ \bibnamefont
  {Cheona}}, \bibinfo {author} {\bibfnamefont {K.-S.}\ \bibnamefont {Koa}},
  \bibinfo {author} {\bibfnamefont {J.-H.}\ \bibnamefont {Choa}}, \bibinfo
  {author} {\bibfnamefont {S.-W.}\ \bibnamefont {Kimb}}, \bibinfo {author}
  {\bibfnamefont {E.-Y.}\ \bibnamefont {Chinc}}, \ and\ \bibinfo {author}
  {\bibfnamefont {H.-T.}\ \bibnamefont {Kimd}},\ }\href@noop {} {\bibfield
  {journal} {\bibinfo  {journal} {J. Electrochem. Soc.}\ }\textbf {\bibinfo
  {volume} {150}},\ \bibinfo {pages} {A796} (\bibinfo {year}
  {2003})}\BibitemShut {NoStop}%
\bibitem [{\citenamefont {Dominko}\ \emph {et~al.}(2011)\citenamefont
  {Dominko}, \citenamefont {Demir-Cakan}, \citenamefont {Morcrette},\ and\
  \citenamefont {Tarascon}}]{dominko2011}%
  \BibitemOpen
  \bibfield  {author} {\bibinfo {author} {\bibfnamefont {R.}~\bibnamefont
  {Dominko}}, \bibinfo {author} {\bibfnamefont {R.}~\bibnamefont
  {Demir-Cakan}}, \bibinfo {author} {\bibfnamefont {M.}~\bibnamefont
  {Morcrette}}, \ and\ \bibinfo {author} {\bibfnamefont {J.-M.}\ \bibnamefont
  {Tarascon}},\ }\href@noop {} {\bibfield  {journal} {\bibinfo  {journal}
  {Electrochemistry Communications}\ }\textbf {\bibinfo {volume} {13}},\
  \bibinfo {pages} {117} (\bibinfo {year} {2011})}\BibitemShut {NoStop}%
\bibitem [{\citenamefont {Kumaresan}\ \emph {et~al.}(2008)\citenamefont
  {Kumaresan}, \citenamefont {Mikhaylik},\ and\ \citenamefont
  {White}}]{kumaresan2008}%
  \BibitemOpen
  \bibfield  {author} {\bibinfo {author} {\bibfnamefont {K.}~\bibnamefont
  {Kumaresan}}, \bibinfo {author} {\bibfnamefont {Y.}~\bibnamefont
  {Mikhaylik}}, \ and\ \bibinfo {author} {\bibfnamefont {R.~E.}\ \bibnamefont
  {White}},\ }\href@noop {} {\bibfield  {journal} {\bibinfo  {journal} {Journal
  of The Electrochemical Society}\ }\textbf {\bibinfo {volume} {155}},\
  \bibinfo {pages} {A576} (\bibinfo {year} {2008})}\BibitemShut {NoStop}%
\bibitem [{\citenamefont {Lu}\ \emph {et~al.}(2013)\citenamefont {Lu},
  \citenamefont {Gallant}, \citenamefont {Kwabi}, \citenamefont {Harding},
  \citenamefont {Mitchell}, \citenamefont {Whittingham},\ and\ \citenamefont
  {Shao-Horn}}]{lu2012}%
  \BibitemOpen
  \bibfield  {author} {\bibinfo {author} {\bibfnamefont {Y.-C.}\ \bibnamefont
  {Lu}}, \bibinfo {author} {\bibfnamefont {B.~M.}\ \bibnamefont {Gallant}},
  \bibinfo {author} {\bibfnamefont {D.~G.}\ \bibnamefont {Kwabi}}, \bibinfo
  {author} {\bibfnamefont {J.~R.}\ \bibnamefont {Harding}}, \bibinfo {author}
  {\bibfnamefont {R.~R.}\ \bibnamefont {Mitchell}}, \bibinfo {author}
  {\bibfnamefont {M.~S.}\ \bibnamefont {Whittingham}}, \ and\ \bibinfo {author}
  {\bibfnamefont {Y.}~\bibnamefont {Shao-Horn}},\ }\href@noop {} {\bibfield
  {journal} {\bibinfo  {journal} {Energy \& Environmental Science}\ }\textbf
  {\bibinfo {volume} {6}},\ \bibinfo {pages} {750} (\bibinfo {year}
  {2013})}\BibitemShut {NoStop}%
\bibitem [{\citenamefont {Gallant}\ \emph {et~al.}(2012)\citenamefont
  {Gallant}, \citenamefont {Mitchell}, \citenamefont {Kwabi}, \citenamefont
  {Zhou}, \citenamefont {Zuin}, \citenamefont {Thompson},\ and\ \citenamefont
  {Shao-Horn}}]{gallant2012a}%
  \BibitemOpen
  \bibfield  {author} {\bibinfo {author} {\bibfnamefont {B.~M.}\ \bibnamefont
  {Gallant}}, \bibinfo {author} {\bibfnamefont {R.~R.}\ \bibnamefont
  {Mitchell}}, \bibinfo {author} {\bibfnamefont {D.~G.}\ \bibnamefont {Kwabi}},
  \bibinfo {author} {\bibfnamefont {J.}~\bibnamefont {Zhou}}, \bibinfo {author}
  {\bibfnamefont {L.}~\bibnamefont {Zuin}}, \bibinfo {author} {\bibfnamefont
  {C.~V.}\ \bibnamefont {Thompson}}, \ and\ \bibinfo {author} {\bibfnamefont
  {Y.}~\bibnamefont {Shao-Horn}},\ }\href@noop {} {\bibfield  {journal}
  {\bibinfo  {journal} {The Journal of Physical Chemistry C}\ }\textbf
  {\bibinfo {volume} {116}},\ \bibinfo {pages} {20800} (\bibinfo {year}
  {2012})}\BibitemShut {NoStop}%
\bibitem [{\citenamefont {Ramadesigan}\ \emph {et~al.}(2012)\citenamefont
  {Ramadesigan}, \citenamefont {Northrop}, \citenamefont {De}, \citenamefont
  {Santhanagopalan}, \citenamefont {Braatz},\ and\ \citenamefont
  {Subramanian}}]{ramadesigan2012}%
  \BibitemOpen
  \bibfield  {author} {\bibinfo {author} {\bibfnamefont {V.}~\bibnamefont
  {Ramadesigan}}, \bibinfo {author} {\bibfnamefont {P.~W.~C.}\ \bibnamefont
  {Northrop}}, \bibinfo {author} {\bibfnamefont {S.}~\bibnamefont {De}},
  \bibinfo {author} {\bibfnamefont {S.}~\bibnamefont {Santhanagopalan}},
  \bibinfo {author} {\bibfnamefont {R.~D.}\ \bibnamefont {Braatz}}, \ and\
  \bibinfo {author} {\bibfnamefont {V.~R.}\ \bibnamefont {Subramanian}},\
  }\href@noop {} {\bibfield  {journal} {\bibinfo  {journal} {Journal of The
  Electrochemical Society}\ }\textbf {\bibinfo {volume} {159}},\ \bibinfo
  {pages} {R31} (\bibinfo {year} {2012})}\BibitemShut {NoStop}%
\bibitem [{\citenamefont {Safari}\ and\ \citenamefont
  {Delacourt}(2009)}]{safari2009}%
  \BibitemOpen
  \bibfield  {author} {\bibinfo {author} {\bibfnamefont {M.}~\bibnamefont
  {Safari}}\ and\ \bibinfo {author} {\bibfnamefont {C.}~\bibnamefont
  {Delacourt}},\ }\href@noop {} {\bibfield  {journal} {\bibinfo  {journal} {J.
  Electrochem. Soc. 2011}\ }\textbf {\bibinfo {volume} {158}},\ \bibinfo
  {pages} {A562} (\bibinfo {year} {2009})}\BibitemShut {NoStop}%
\bibitem [{\citenamefont {Delacourt}\ and\ \citenamefont
  {Safari}(2011)}]{delacourt2011}%
  \BibitemOpen
  \bibfield  {author} {\bibinfo {author} {\bibfnamefont {C.}~\bibnamefont
  {Delacourt}}\ and\ \bibinfo {author} {\bibfnamefont {M.}~\bibnamefont
  {Safari}},\ }\href@noop {} {\bibfield  {journal} {\bibinfo  {journal}
  {Electrochimica Acta}\ }\textbf {\bibinfo {volume} {56}},\ \bibinfo {pages}
  {5222} (\bibinfo {year} {2011})}\BibitemShut {NoStop}%
\bibitem [{\citenamefont {Srinivasan}\ and\ \citenamefont
  {Newman}(2004)}]{srinivasan2004}%
  \BibitemOpen
  \bibfield  {author} {\bibinfo {author} {\bibfnamefont {V.}~\bibnamefont
  {Srinivasan}}\ and\ \bibinfo {author} {\bibfnamefont {J.}~\bibnamefont
  {Newman}},\ }\href@noop {} {\bibfield  {journal} {\bibinfo  {journal}
  {Journal of the Electrochemical Society}\ }\textbf {\bibinfo {volume}
  {151}},\ \bibinfo {pages} {A1517} (\bibinfo {year} {2004})}\BibitemShut
  {NoStop}%
\bibitem [{\citenamefont {Newman}\ and\ \citenamefont
  {Tiedemann}(1975)}]{newman1975}%
  \BibitemOpen
  \bibfield  {author} {\bibinfo {author} {\bibfnamefont {J.}~\bibnamefont
  {Newman}}\ and\ \bibinfo {author} {\bibfnamefont {W.}~\bibnamefont
  {Tiedemann}},\ }\href@noop {} {\bibfield  {journal} {\bibinfo  {journal}
  {AIChE Journal}\ }\textbf {\bibinfo {volume} {21}},\ \bibinfo {pages} {25}
  (\bibinfo {year} {1975})}\BibitemShut {NoStop}%
\bibitem [{\citenamefont {Newman}\ and\ \citenamefont
  {Thomas-Alyea}(2004)}]{newman_book}%
  \BibitemOpen
  \bibfield  {author} {\bibinfo {author} {\bibfnamefont {J.}~\bibnamefont
  {Newman}}\ and\ \bibinfo {author} {\bibfnamefont {K.~E.}\ \bibnamefont
  {Thomas-Alyea}},\ }\href@noop {} {\emph {\bibinfo {title} {Electrochemical
  Systems}}},\ \bibinfo {edition} {3rd}\ ed.\ (\bibinfo  {publisher}
  {Prentice-Hall, Inc.},\ \bibinfo {address} {Englewood Cliffs, NJ},\ \bibinfo
  {year} {2004})\BibitemShut {NoStop}%
\bibitem [{\citenamefont {Lai}\ and\ \citenamefont {Ciucci}(2010)}]{lai2010}%
  \BibitemOpen
  \bibfield  {author} {\bibinfo {author} {\bibfnamefont {W.}~\bibnamefont
  {Lai}}\ and\ \bibinfo {author} {\bibfnamefont {F.}~\bibnamefont {Ciucci}},\
  }\href@noop {} {\bibfield  {journal} {\bibinfo  {journal} {Electrochim.
  Acta}\ }\textbf {\bibinfo {volume} {56}},\ \bibinfo {pages} {531} (\bibinfo
  {year} {2010})}\BibitemShut {NoStop}%
\bibitem [{\citenamefont {Lai}\ and\ \citenamefont {Ciucci}(2011)}]{lai2011a}%
  \BibitemOpen
  \bibfield  {author} {\bibinfo {author} {\bibfnamefont {W.}~\bibnamefont
  {Lai}}\ and\ \bibinfo {author} {\bibfnamefont {F.}~\bibnamefont {Ciucci}},\
  }\href@noop {} {\bibfield  {journal} {\bibinfo  {journal} {Electrochimica
  Acta}\ }\textbf {\bibinfo {volume} {56}},\ \bibinfo {pages} {4369} (\bibinfo
  {year} {2011})}\BibitemShut {NoStop}%
\bibitem [{\citenamefont {Ferguson}\ and\ \citenamefont
  {Bazant}(2012)}]{ferguson2012}%
  \BibitemOpen
  \bibfield  {author} {\bibinfo {author} {\bibfnamefont {T.~R.}\ \bibnamefont
  {Ferguson}}\ and\ \bibinfo {author} {\bibfnamefont {M.~Z.}\ \bibnamefont
  {Bazant}},\ }\href {\doibase 10.1149/2.048212jes} {\bibfield  {journal}
  {\bibinfo  {journal} {J. Electrochem. Soc.}\ }\textbf {\bibinfo {volume}
  {159}},\ \bibinfo {pages} {A1967} (\bibinfo {year} {2012})}\BibitemShut
  {NoStop}%
\bibitem [{\citenamefont {Dargaville}\ and\ \citenamefont
  {Farrell}(2010)}]{dargaville2010}%
  \BibitemOpen
  \bibfield  {author} {\bibinfo {author} {\bibfnamefont {S.}~\bibnamefont
  {Dargaville}}\ and\ \bibinfo {author} {\bibfnamefont {T.}~\bibnamefont
  {Farrell}},\ }\href@noop {} {\bibfield  {journal} {\bibinfo  {journal}
  {Journal of the Electrochemical Society}\ }\textbf {\bibinfo {volume}
  {157}},\ \bibinfo {pages} {A830} (\bibinfo {year} {2010})}\BibitemShut
  {NoStop}%
\bibitem [{\citenamefont {Wang}\ \emph
  {et~al.}(2007{\natexlab{a}})\citenamefont {Wang}, \citenamefont
  {Kasavajjula},\ and\ \citenamefont {Arce}}]{cwang2007}%
  \BibitemOpen
  \bibfield  {author} {\bibinfo {author} {\bibfnamefont {C.}~\bibnamefont
  {Wang}}, \bibinfo {author} {\bibfnamefont {U.~S.}\ \bibnamefont
  {Kasavajjula}}, \ and\ \bibinfo {author} {\bibfnamefont {P.~E.}\ \bibnamefont
  {Arce}},\ }\href@noop {} {\bibfield  {journal} {\bibinfo  {journal} {J. Phys.
  Chem. C}\ }\textbf {\bibinfo {volume} {111}},\ \bibinfo {pages} {16656}
  (\bibinfo {year} {2007}{\natexlab{a}})}\BibitemShut {NoStop}%
\bibitem [{\citenamefont {Kasavajjula}\ \emph {et~al.}(2008)\citenamefont
  {Kasavajjula}, \citenamefont {Wang},\ and\ \citenamefont
  {Arce}}]{kasavajjula2008}%
  \BibitemOpen
  \bibfield  {author} {\bibinfo {author} {\bibfnamefont {U.~S.}\ \bibnamefont
  {Kasavajjula}}, \bibinfo {author} {\bibfnamefont {C.}~\bibnamefont {Wang}}, \
  and\ \bibinfo {author} {\bibfnamefont {P.~E.}\ \bibnamefont {Arce}},\
  }\href@noop {} {\bibfield  {journal} {\bibinfo  {journal} {J. Electrochem.
  Soc.}\ }\textbf {\bibinfo {volume} {155}},\ \bibinfo {pages} {A866} (\bibinfo
  {year} {2008})}\BibitemShut {NoStop}%
\bibitem [{\citenamefont {Zhu}\ and\ \citenamefont {Wang}(2010)}]{zhu2010}%
  \BibitemOpen
  \bibfield  {author} {\bibinfo {author} {\bibfnamefont {Y.}~\bibnamefont
  {Zhu}}\ and\ \bibinfo {author} {\bibfnamefont {C.}~\bibnamefont {Wang}},\
  }\href@noop {} {\bibfield  {journal} {\bibinfo  {journal} {Journal of
  Physical Chemistry C}\ }\textbf {\bibinfo {volume} {114}},\ \bibinfo {pages}
  {2830} (\bibinfo {year} {2010})}\BibitemShut {NoStop}%
\bibitem [{\citenamefont {der Ven}\ \emph {et~al.}(2009)\citenamefont {der
  Ven}, \citenamefont {Garikipati}, \citenamefont {Kim},\ and\ \citenamefont
  {Wagemaker}}]{vanderven2009}%
  \BibitemOpen
  \bibfield  {author} {\bibinfo {author} {\bibfnamefont {A.~V.}\ \bibnamefont
  {der Ven}}, \bibinfo {author} {\bibfnamefont {K.}~\bibnamefont {Garikipati}},
  \bibinfo {author} {\bibfnamefont {S.}~\bibnamefont {Kim}}, \ and\ \bibinfo
  {author} {\bibfnamefont {M.}~\bibnamefont {Wagemaker}},\ }\href@noop {}
  {\bibfield  {journal} {\bibinfo  {journal} {J. Electrochem. Soc.}\ }\textbf
  {\bibinfo {volume} {156}},\ \bibinfo {pages} {A949} (\bibinfo {year}
  {2009})}\BibitemShut {NoStop}%
\bibitem [{\citenamefont {Bai}\ \emph {et~al.}(2011)\citenamefont {Bai},
  \citenamefont {Cogswell},\ and\ \citenamefont {Bazant}}]{bai2011}%
  \BibitemOpen
  \bibfield  {author} {\bibinfo {author} {\bibfnamefont {P.}~\bibnamefont
  {Bai}}, \bibinfo {author} {\bibfnamefont {D.~A.}\ \bibnamefont {Cogswell}}, \
  and\ \bibinfo {author} {\bibfnamefont {M.~Z.}\ \bibnamefont {Bazant}},\
  }\href@noop {} {\bibfield  {journal} {\bibinfo  {journal} {Nano Letters}\
  }\textbf {\bibinfo {volume} {11}},\ \bibinfo {pages} {4890} (\bibinfo {year}
  {2011})}\BibitemShut {NoStop}%
\bibitem [{\citenamefont {Cogswell}\ and\ \citenamefont
  {Bazant}(2012)}]{cogswell2012}%
  \BibitemOpen
  \bibfield  {author} {\bibinfo {author} {\bibfnamefont {D.~A.}\ \bibnamefont
  {Cogswell}}\ and\ \bibinfo {author} {\bibfnamefont {M.~Z.}\ \bibnamefont
  {Bazant}},\ }\href {\doibase 10.1021/nn204177u} {\bibfield  {journal}
  {\bibinfo  {journal} {ACS Nano}\ }\textbf {\bibinfo {volume} {6}},\ \bibinfo
  {pages} {2215} (\bibinfo {year} {2012})}\BibitemShut {NoStop}%
\bibitem [{\citenamefont {Lai}(2011)}]{lai2011b}%
  \BibitemOpen
  \bibfield  {author} {\bibinfo {author} {\bibfnamefont {W.}~\bibnamefont
  {Lai}},\ }\href@noop {} {\bibfield  {journal} {\bibinfo  {journal} {Journal
  of Power Sources}\ }\textbf {\bibinfo {volume} {196}},\ \bibinfo {pages}
  {6534} (\bibinfo {year} {2011})}\BibitemShut {NoStop}%
\bibitem [{\citenamefont {Dargaville}\ and\ \citenamefont
  {Farrell}(2013{\natexlab{a}})}]{dargaville2013CHR}%
  \BibitemOpen
  \bibfield  {author} {\bibinfo {author} {\bibfnamefont {S.}~\bibnamefont
  {Dargaville}}\ and\ \bibinfo {author} {\bibfnamefont {T.~W.}\ \bibnamefont
  {Farrell}},\ }\href@noop {} {\bibfield  {journal} {\bibinfo  {journal}
  {Electrochimica Acta}\ }\textbf {\bibinfo {volume} {94}},\ \bibinfo {pages}
  {143} (\bibinfo {year} {2013}{\natexlab{a}})}\BibitemShut {NoStop}%
\bibitem [{\citenamefont {Zeng}\ and\ \citenamefont
  {Bazant}(2013{\natexlab{a}})}]{zeng2013MRS}%
  \BibitemOpen
  \bibfield  {author} {\bibinfo {author} {\bibfnamefont {Y.}~\bibnamefont
  {Zeng}}\ and\ \bibinfo {author} {\bibfnamefont {M.~Z.}\ \bibnamefont
  {Bazant}},\ }\href@noop {} {\bibfield  {journal} {\bibinfo  {journal} {MRS
  Proceedings}\ }\textbf {\bibinfo {volume} {1542}} (\bibinfo {year}
  {2013}{\natexlab{a}})}\BibitemShut {NoStop}%
\bibitem [{\citenamefont {Zeng}\ and\ \citenamefont
  {Bazant}(2013{\natexlab{b}})}]{zeng2014}%
  \BibitemOpen
  \bibfield  {author} {\bibinfo {author} {\bibfnamefont {Y.}~\bibnamefont
  {Zeng}}\ and\ \bibinfo {author} {\bibfnamefont {M.~Z.}\ \bibnamefont
  {Bazant}},\ }\href@noop {} {\bibfield  {journal} {\bibinfo  {journal}
  {submitted}\ }\textbf {\bibinfo {volume} {arXiv:1309.4543 [physics.chem-ph]}}
  (\bibinfo {year} {2013}{\natexlab{b}})}\BibitemShut {NoStop}%
\bibitem [{\citenamefont {Dreyer}\ \emph {et~al.}(2010)\citenamefont {Dreyer},
  \citenamefont {Jamnik}, \citenamefont {Guhlke}, \citenamefont {Huth},
  \citenamefont {Moskon},\ and\ \citenamefont {Gaberscek}}]{dreyer2010}%
  \BibitemOpen
  \bibfield  {author} {\bibinfo {author} {\bibfnamefont {W.}~\bibnamefont
  {Dreyer}}, \bibinfo {author} {\bibfnamefont {J.}~\bibnamefont {Jamnik}},
  \bibinfo {author} {\bibfnamefont {C.}~\bibnamefont {Guhlke}}, \bibinfo
  {author} {\bibfnamefont {R.}~\bibnamefont {Huth}}, \bibinfo {author}
  {\bibfnamefont {J.}~\bibnamefont {Moskon}}, \ and\ \bibinfo {author}
  {\bibfnamefont {M.}~\bibnamefont {Gaberscek}},\ }\href {\doibase
  10.1038/nmat2730} {\bibfield  {journal} {\bibinfo  {journal} {Nat. Mater.}\
  }\textbf {\bibinfo {volume} {9}},\ \bibinfo {pages} {448} (\bibinfo {year}
  {2010})}\BibitemShut {NoStop}%
\bibitem [{\citenamefont {Dreyer}\ \emph {et~al.}(2011)\citenamefont {Dreyer},
  \citenamefont {Guhlke},\ and\ \citenamefont {Huth}}]{dreyer2011}%
  \BibitemOpen
  \bibfield  {author} {\bibinfo {author} {\bibfnamefont {D.}~\bibnamefont
  {Dreyer}}, \bibinfo {author} {\bibfnamefont {C.}~\bibnamefont {Guhlke}}, \
  and\ \bibinfo {author} {\bibfnamefont {R.}~\bibnamefont {Huth}},\ }\href
  {\doibase 10.1016/j.physd.2011.02.011} {\bibfield  {journal} {\bibinfo
  {journal} {Physica D}\ }\textbf {\bibinfo {volume} {240}},\ \bibinfo {pages}
  {1008} (\bibinfo {year} {2011})}\BibitemShut {NoStop}%
\bibitem [{\citenamefont {Bai}\ and\ \citenamefont {Tian}(2013)}]{bai2013}%
  \BibitemOpen
  \bibfield  {author} {\bibinfo {author} {\bibfnamefont {P.}~\bibnamefont
  {Bai}}\ and\ \bibinfo {author} {\bibfnamefont {G.}~\bibnamefont {Tian}},\
  }\href@noop {} {\bibfield  {journal} {\bibinfo  {journal} {Electrochimica
  Acta}\ }\textbf {\bibinfo {volume} {89}},\ \bibinfo {pages} {644} (\bibinfo
  {year} {2013})}\BibitemShut {NoStop}%
\bibitem [{\citenamefont {Levi}\ \emph {et~al.}(2013)\citenamefont {Levi},
  \citenamefont {Sigalov}, \citenamefont {Salitra}, \citenamefont {Nayak},
  \citenamefont {Aurbach}, \citenamefont {Daikhin}, \citenamefont {Perre},\
  and\ \citenamefont {Presser}}]{levi2013}%
  \BibitemOpen
  \bibfield  {author} {\bibinfo {author} {\bibfnamefont {M.~D.}\ \bibnamefont
  {Levi}}, \bibinfo {author} {\bibfnamefont {S.}~\bibnamefont {Sigalov}},
  \bibinfo {author} {\bibfnamefont {G.}~\bibnamefont {Salitra}}, \bibinfo
  {author} {\bibfnamefont {P.}~\bibnamefont {Nayak}}, \bibinfo {author}
  {\bibfnamefont {D.}~\bibnamefont {Aurbach}}, \bibinfo {author} {\bibfnamefont
  {L.}~\bibnamefont {Daikhin}}, \bibinfo {author} {\bibfnamefont
  {E.}~\bibnamefont {Perre}}, \ and\ \bibinfo {author} {\bibfnamefont
  {V.}~\bibnamefont {Presser}},\ }\href@noop {} {\bibfield  {journal} {\bibinfo
   {journal} {J. Phys. Chem. C}\ }\textbf {\bibinfo {volume} {117}},\ \bibinfo
  {pages} {15505} (\bibinfo {year} {2013})}\BibitemShut {NoStop}%
\bibitem [{\citenamefont {Meethong}\ \emph {et~al.}(2007)\citenamefont
  {Meethong}, \citenamefont {Huang}, \citenamefont {Carter},\ and\
  \citenamefont {Chiang}}]{meethong2007}%
  \BibitemOpen
  \bibfield  {author} {\bibinfo {author} {\bibfnamefont {N.}~\bibnamefont
  {Meethong}}, \bibinfo {author} {\bibfnamefont {H.-Y.~S.}\ \bibnamefont
  {Huang}}, \bibinfo {author} {\bibfnamefont {W.~C.}\ \bibnamefont {Carter}}, \
  and\ \bibinfo {author} {\bibfnamefont {Y.-M.}\ \bibnamefont {Chiang}},\
  }\href@noop {} {\bibfield  {journal} {\bibinfo  {journal} {Electrochem.
  Solid-State Lett.}\ }\textbf {\bibinfo {volume} {10}},\ \bibinfo {pages}
  {A134} (\bibinfo {year} {2007})}\BibitemShut {NoStop}%
\bibitem [{\citenamefont {Tang}\ \emph {et~al.}(2009)\citenamefont {Tang},
  \citenamefont {Huang}, \citenamefont {Meethong}, \citenamefont {Kao},
  \citenamefont {Carter},\ and\ \citenamefont {Chiang}}]{tang2009}%
  \BibitemOpen
  \bibfield  {author} {\bibinfo {author} {\bibfnamefont {M.}~\bibnamefont
  {Tang}}, \bibinfo {author} {\bibfnamefont {H.-Y.}\ \bibnamefont {Huang}},
  \bibinfo {author} {\bibfnamefont {N.}~\bibnamefont {Meethong}}, \bibinfo
  {author} {\bibfnamefont {Y.-H.}\ \bibnamefont {Kao}}, \bibinfo {author}
  {\bibfnamefont {W.~C.}\ \bibnamefont {Carter}}, \ and\ \bibinfo {author}
  {\bibfnamefont {Y.-M.}\ \bibnamefont {Chiang}},\ }\href@noop {} {\bibfield
  {journal} {\bibinfo  {journal} {Chem. Mater.}\ }\textbf {\bibinfo {volume}
  {21}},\ \bibinfo {pages} {1557} (\bibinfo {year} {2009})}\BibitemShut
  {NoStop}%
\bibitem [{\citenamefont {Cogswell}\ and\ \citenamefont
  {Bazant}(2013)}]{cogswell2013}%
  \BibitemOpen
  \bibfield  {author} {\bibinfo {author} {\bibfnamefont {D.~A.}\ \bibnamefont
  {Cogswell}}\ and\ \bibinfo {author} {\bibfnamefont {M.~Z.}\ \bibnamefont
  {Bazant}},\ }\href {\doibase 10.1021/nl400497t} {\bibfield  {journal}
  {\bibinfo  {journal} {Nano Letters}\ }\textbf {\bibinfo {volume} {13}},\
  \bibinfo {pages} {3036} (\bibinfo {year} {2013})}\BibitemShut {NoStop}%
\bibitem [{\citenamefont {Burch}(2009)}]{burch_thesis}%
  \BibitemOpen
  \bibfield  {author} {\bibinfo {author} {\bibfnamefont {D.}~\bibnamefont
  {Burch}},\ }\href@noop {} {\emph {\bibinfo {title} {Intercalation Dynamics in
  Lithium-Ion Batteries}}}\ (\bibinfo  {publisher} {Ph.D. Thesis in
  Mathematics, Massachusetts Institute of Technology},\ \bibinfo {year}
  {2009})\BibitemShut {NoStop}%
\bibitem [{\citenamefont {Dargaville}\ and\ \citenamefont
  {Farrell}(2013{\natexlab{b}})}]{dargaville2013porous}%
  \BibitemOpen
  \bibfield  {author} {\bibinfo {author} {\bibfnamefont {S.}~\bibnamefont
  {Dargaville}}\ and\ \bibinfo {author} {\bibfnamefont {T.}~\bibnamefont
  {Farrell}},\ }\href@noop {} {\bibfield  {journal} {\bibinfo  {journal}
  {Electrochimica Acta}\ }\textbf {\bibinfo {volume} {111}},\ \bibinfo {pages}
  {474} (\bibinfo {year} {2013}{\natexlab{b}})}\BibitemShut {NoStop}%
\bibitem [{\citenamefont {Garcia}\ \emph {et~al.}(2005)\citenamefont {Garcia},
  \citenamefont {Chiang}, \citenamefont {Carter}, \citenamefont {Limthongkul},\
  and\ \citenamefont {Bishop}}]{garcia2005}%
  \BibitemOpen
  \bibfield  {author} {\bibinfo {author} {\bibfnamefont {R.~E.}\ \bibnamefont
  {Garcia}}, \bibinfo {author} {\bibfnamefont {Y.-M.}\ \bibnamefont {Chiang}},
  \bibinfo {author} {\bibfnamefont {W.~C.}\ \bibnamefont {Carter}}, \bibinfo
  {author} {\bibfnamefont {P.}~\bibnamefont {Limthongkul}}, \ and\ \bibinfo
  {author} {\bibfnamefont {C.~M.}\ \bibnamefont {Bishop}},\ }\href@noop {}
  {\bibfield  {journal} {\bibinfo  {journal} {Journal of the Electrochemical
  Society}\ }\textbf {\bibinfo {volume} {152}},\ \bibinfo {pages} {A255}
  (\bibinfo {year} {2005})}\BibitemShut {NoStop}%
\bibitem [{\citenamefont {Garcia}\ and\ \citenamefont
  {Chiang}(2007)}]{garcia2007}%
  \BibitemOpen
  \bibfield  {author} {\bibinfo {author} {\bibfnamefont {R.~E.}\ \bibnamefont
  {Garcia}}\ and\ \bibinfo {author} {\bibfnamefont {Y.-M.}\ \bibnamefont
  {Chiang}},\ }\href@noop {} {\bibfield  {journal} {\bibinfo  {journal}
  {Journal of the Electrochemical Society}\ }\textbf {\bibinfo {volume}
  {154}},\ \bibinfo {pages} {A856} (\bibinfo {year} {2007})}\BibitemShut
  {NoStop}%
\bibitem [{\citenamefont {Orvananos}\ \emph {et~al.}(2014)\citenamefont
  {Orvananos}, \citenamefont {Ferguson}, \citenamefont {Yu}, \citenamefont
  {Bazant},\ and\ \citenamefont {Thornton}}]{orvananos2013}%
  \BibitemOpen
  \bibfield  {author} {\bibinfo {author} {\bibfnamefont {B.}~\bibnamefont
  {Orvananos}}, \bibinfo {author} {\bibfnamefont {T.~R.}\ \bibnamefont
  {Ferguson}}, \bibinfo {author} {\bibfnamefont {H.-C.}\ \bibnamefont {Yu}},
  \bibinfo {author} {\bibfnamefont {M.~Z.}\ \bibnamefont {Bazant}}, \ and\
  \bibinfo {author} {\bibfnamefont {K.}~\bibnamefont {Thornton}},\ }\href@noop
  {} {\bibfield  {journal} {\bibinfo  {journal} {J. Electrochem. Soc.}\
  }\textbf {\bibinfo {volume} {in press}} (\bibinfo {year} {2014})},\ \bibinfo
  {note} {arXiv:1309.6495 [cond-mat.mtrl-sci]}\BibitemShut {NoStop}%
\bibitem [{\citenamefont {Kao}\ \emph {et~al.}(2010)\citenamefont {Kao},
  \citenamefont {Tang}, \citenamefont {Meethong}, \citenamefont {Bai},
  \citenamefont {Carter},\ and\ \citenamefont {Chiang}}]{kao2010}%
  \BibitemOpen
  \bibfield  {author} {\bibinfo {author} {\bibfnamefont {Y.-H.}\ \bibnamefont
  {Kao}}, \bibinfo {author} {\bibfnamefont {M.}~\bibnamefont {Tang}}, \bibinfo
  {author} {\bibfnamefont {N.}~\bibnamefont {Meethong}}, \bibinfo {author}
  {\bibfnamefont {J.}~\bibnamefont {Bai}}, \bibinfo {author} {\bibfnamefont
  {W.~C.}\ \bibnamefont {Carter}}, \ and\ \bibinfo {author} {\bibfnamefont
  {Y.-M.}\ \bibnamefont {Chiang}},\ }\href@noop {} {\bibfield  {journal}
  {\bibinfo  {journal} {Chem. Mater.}\ }\textbf {\bibinfo {volume} {22}},\
  \bibinfo {pages} {5845} (\bibinfo {year} {2010})}\BibitemShut {NoStop}%
\bibitem [{\citenamefont {Islam}\ \emph {et~al.}(2005)\citenamefont {Islam},
  \citenamefont {Driscoll}, \citenamefont {Fisher},\ and\ \citenamefont
  {Slater}}]{islam2005}%
  \BibitemOpen
  \bibfield  {author} {\bibinfo {author} {\bibfnamefont {M.~S.}\ \bibnamefont
  {Islam}}, \bibinfo {author} {\bibfnamefont {D.~J.}\ \bibnamefont {Driscoll}},
  \bibinfo {author} {\bibfnamefont {C.~A.~J.}\ \bibnamefont {Fisher}}, \ and\
  \bibinfo {author} {\bibfnamefont {P.~R.}\ \bibnamefont {Slater}},\
  }\href@noop {} {\bibfield  {journal} {\bibinfo  {journal} {Chem. Mater.}\
  }\textbf {\bibinfo {volume} {17}},\ \bibinfo {pages} {5085} (\bibinfo {year}
  {2005})}\BibitemShut {NoStop}%
\bibitem [{\citenamefont {Singh}\ \emph {et~al.}(2008)\citenamefont {Singh},
  \citenamefont {Burch},\ and\ \citenamefont {Bazant}}]{singh2008}%
  \BibitemOpen
  \bibfield  {author} {\bibinfo {author} {\bibfnamefont {G.}~\bibnamefont
  {Singh}}, \bibinfo {author} {\bibfnamefont {D.}~\bibnamefont {Burch}}, \ and\
  \bibinfo {author} {\bibfnamefont {M.~Z.}\ \bibnamefont {Bazant}},\
  }\href@noop {} {\bibfield  {journal} {\bibinfo  {journal} {Electrochimica
  Acta}\ }\textbf {\bibinfo {volume} {53}},\ \bibinfo {pages} {7599} (\bibinfo
  {year} {2008})},\ \bibinfo {note} {arXiv:0707.1858v1 [cond-mat.mtrl-sci]
  (2007)}\BibitemShut {NoStop}%
\bibitem [{\citenamefont {Delmas}\ \emph {et~al.}(2008)\citenamefont {Delmas},
  \citenamefont {Maccario}, \citenamefont {Croguennec}, \citenamefont {Cras},\
  and\ \citenamefont {Weill}}]{delmas2008}%
  \BibitemOpen
  \bibfield  {author} {\bibinfo {author} {\bibfnamefont {C.}~\bibnamefont
  {Delmas}}, \bibinfo {author} {\bibfnamefont {M.}~\bibnamefont {Maccario}},
  \bibinfo {author} {\bibfnamefont {L.}~\bibnamefont {Croguennec}}, \bibinfo
  {author} {\bibfnamefont {F.~L.}\ \bibnamefont {Cras}}, \ and\ \bibinfo
  {author} {\bibfnamefont {F.}~\bibnamefont {Weill}},\ }\href@noop {}
  {\bibfield  {journal} {\bibinfo  {journal} {Nature Materials}\ }\textbf
  {\bibinfo {volume} {7}},\ \bibinfo {pages} {665} (\bibinfo {year}
  {2008})}\BibitemShut {NoStop}%
\bibitem [{\citenamefont {Tang}\ \emph {et~al.}(2011)\citenamefont {Tang},
  \citenamefont {Belak},\ and\ \citenamefont {Dorr}}]{tang2011}%
  \BibitemOpen
  \bibfield  {author} {\bibinfo {author} {\bibfnamefont {M.}~\bibnamefont
  {Tang}}, \bibinfo {author} {\bibfnamefont {J.~F.}\ \bibnamefont {Belak}}, \
  and\ \bibinfo {author} {\bibfnamefont {M.~R.}\ \bibnamefont {Dorr}},\
  }\href@noop {} {\bibfield  {journal} {\bibinfo  {journal} {The Journal of
  Physical Chemistry C}\ }\textbf {\bibinfo {volume} {115}},\ \bibinfo {pages}
  {4922} (\bibinfo {year} {2011})}\BibitemShut {NoStop}%
\bibitem [{\citenamefont {Malik}\ \emph {et~al.}(2011)\citenamefont {Malik},
  \citenamefont {Zhou},\ and\ \citenamefont {Ceder}}]{malik2011}%
  \BibitemOpen
  \bibfield  {author} {\bibinfo {author} {\bibfnamefont {R.}~\bibnamefont
  {Malik}}, \bibinfo {author} {\bibfnamefont {F.}~\bibnamefont {Zhou}}, \ and\
  \bibinfo {author} {\bibfnamefont {G.}~\bibnamefont {Ceder}},\ }\href@noop {}
  {\bibfield  {journal} {\bibinfo  {journal} {Nature Materials}\ }\textbf
  {\bibinfo {volume} {10}},\ \bibinfo {pages} {587} (\bibinfo {year}
  {2011})}\BibitemShut {NoStop}%
\bibitem [{\citenamefont {Ichitsubo}\ \emph {et~al.}(2013)\citenamefont
  {Ichitsubo}, \citenamefont {Doi}, \citenamefont {Tokuda}, \citenamefont
  {Matsubara}, \citenamefont {Kida}, \citenamefont {Kawaguchi}, \citenamefont
  {Yagi}, \citenamefont {Okada},\ and\ \citenamefont {Yamaki}}]{ichitsubo2013}%
  \BibitemOpen
  \bibfield  {author} {\bibinfo {author} {\bibfnamefont {T.}~\bibnamefont
  {Ichitsubo}}, \bibinfo {author} {\bibfnamefont {T.}~\bibnamefont {Doi}},
  \bibinfo {author} {\bibfnamefont {K.}~\bibnamefont {Tokuda}}, \bibinfo
  {author} {\bibfnamefont {E.}~\bibnamefont {Matsubara}}, \bibinfo {author}
  {\bibfnamefont {T.}~\bibnamefont {Kida}}, \bibinfo {author} {\bibfnamefont
  {T.}~\bibnamefont {Kawaguchi}}, \bibinfo {author} {\bibfnamefont
  {S.}~\bibnamefont {Yagi}}, \bibinfo {author} {\bibfnamefont {S.}~\bibnamefont
  {Okada}}, \ and\ \bibinfo {author} {\bibfnamefont {J.}~\bibnamefont
  {Yamaki}},\ }\href@noop {} {\bibfield  {journal} {\bibinfo  {journal} {J.
  Mater. Chem. A}\ }\textbf {\bibinfo {volume} {1}},\ \bibinfo {pages} {14532}
  (\bibinfo {year} {2013})}\BibitemShut {NoStop}%
\bibitem [{\citenamefont {Burch}\ and\ \citenamefont
  {Bazant}(2009)}]{burch2009}%
  \BibitemOpen
  \bibfield  {author} {\bibinfo {author} {\bibfnamefont {D.}~\bibnamefont
  {Burch}}\ and\ \bibinfo {author} {\bibfnamefont {M.~Z.}\ \bibnamefont
  {Bazant}},\ }\href@noop {} {\bibfield  {journal} {\bibinfo  {journal} {Nano
  Letters}\ }\textbf {\bibinfo {volume} {9}},\ \bibinfo {pages} {3795}
  (\bibinfo {year} {2009})}\BibitemShut {NoStop}%
\bibitem [{\citenamefont {Oyama}\ \emph {et~al.}(2012)\citenamefont {Oyama},
  \citenamefont {Yamada}, \citenamefont {Natsui}, \citenamefont {Nishimura},\
  and\ \citenamefont {Yamada}}]{oyama2012}%
  \BibitemOpen
  \bibfield  {author} {\bibinfo {author} {\bibfnamefont {G.}~\bibnamefont
  {Oyama}}, \bibinfo {author} {\bibfnamefont {Y.}~\bibnamefont {Yamada}},
  \bibinfo {author} {\bibfnamefont {R.}~\bibnamefont {Natsui}}, \bibinfo
  {author} {\bibfnamefont {S.}~\bibnamefont {Nishimura}}, \ and\ \bibinfo
  {author} {\bibfnamefont {A.}~\bibnamefont {Yamada}},\ }\href@noop {}
  {\bibfield  {journal} {\bibinfo  {journal} {J. Phys. Chem. C}\ }\textbf
  {\bibinfo {volume} {116}},\ \bibinfo {pages} {7306} (\bibinfo {year}
  {2012})}\BibitemShut {NoStop}%
\bibitem [{\citenamefont {Bai}\ and\ \citenamefont {Bazant}()}]{bai2014}%
  \BibitemOpen
  \bibfield  {author} {\bibinfo {author} {\bibfnamefont {P.}~\bibnamefont
  {Bai}}\ and\ \bibinfo {author} {\bibfnamefont {M.~Z.}\ \bibnamefont
  {Bazant}},\ }\href@noop {} {\enquote {\bibinfo {title} {Charge transfer
  kinetics at the solid-solid interface in porous electrodes},}\ }\BibitemShut
  {NoStop}%
\bibitem [{\citenamefont {Smith}\ \emph {et~al.}(2012)\citenamefont {Smith},
  \citenamefont {Mukherjee},\ and\ \citenamefont {Fisher}}]{smith2012}%
  \BibitemOpen
  \bibfield  {author} {\bibinfo {author} {\bibfnamefont {K.~C.}\ \bibnamefont
  {Smith}}, \bibinfo {author} {\bibfnamefont {P.~P.}\ \bibnamefont
  {Mukherjee}}, \ and\ \bibinfo {author} {\bibfnamefont {T.~S.}\ \bibnamefont
  {Fisher}},\ }\href {\doibase 10.1039/C2CP40135E} {\bibfield  {journal}
  {\bibinfo  {journal} {Phys. Chem. Chem. Phys.}\ }\textbf {\bibinfo {volume}
  {14}},\ \bibinfo {pages} {7040} (\bibinfo {year} {2012})}\BibitemShut
  {NoStop}%
\bibitem [{\citenamefont {Malik}\ \emph {et~al.}(2010)\citenamefont {Malik},
  \citenamefont {Burch}, \citenamefont {Bazant},\ and\ \citenamefont
  {Ceder}}]{malik2010}%
  \BibitemOpen
  \bibfield  {author} {\bibinfo {author} {\bibfnamefont {R.}~\bibnamefont
  {Malik}}, \bibinfo {author} {\bibfnamefont {D.}~\bibnamefont {Burch}},
  \bibinfo {author} {\bibfnamefont {M.}~\bibnamefont {Bazant}}, \ and\ \bibinfo
  {author} {\bibfnamefont {G.}~\bibnamefont {Ceder}},\ }\href@noop {}
  {\bibfield  {journal} {\bibinfo  {journal} {Nano Letters}\ }\textbf {\bibinfo
  {volume} {10}},\ \bibinfo {pages} {4123} (\bibinfo {year}
  {2010})}\BibitemShut {NoStop}%
\bibitem [{\citenamefont {Wang}\ \emph
  {et~al.}(2007{\natexlab{b}})\citenamefont {Wang}, \citenamefont {Zhou},
  \citenamefont {Meng},\ and\ \citenamefont {Ceder}}]{wang2007}%
  \BibitemOpen
  \bibfield  {author} {\bibinfo {author} {\bibfnamefont {L.}~\bibnamefont
  {Wang}}, \bibinfo {author} {\bibfnamefont {F.}~\bibnamefont {Zhou}}, \bibinfo
  {author} {\bibfnamefont {Y.~S.}\ \bibnamefont {Meng}}, \ and\ \bibinfo
  {author} {\bibfnamefont {G.}~\bibnamefont {Ceder}},\ }\href@noop {}
  {\bibfield  {journal} {\bibinfo  {journal} {Phys. Rev. B}\ }\textbf {\bibinfo
  {volume} {76}},\ \bibinfo {pages} {165435} (\bibinfo {year}
  {2007}{\natexlab{b}})}\BibitemShut {NoStop}%
\bibitem [{\citenamefont {Maxisch}\ and\ \citenamefont
  {Ceder}(2006)}]{maxisch2006}%
  \BibitemOpen
  \bibfield  {author} {\bibinfo {author} {\bibfnamefont {T.}~\bibnamefont
  {Maxisch}}\ and\ \bibinfo {author} {\bibfnamefont {G.}~\bibnamefont
  {Ceder}},\ }\href@noop {} {\bibfield  {journal} {\bibinfo  {journal} {Phys.
  Rev. B}\ }\textbf {\bibinfo {volume} {73}},\ \bibinfo {pages} {174112}
  (\bibinfo {year} {2006})}\BibitemShut {NoStop}%
\bibitem [{\citenamefont {Bard}\ and\ \citenamefont
  {Faulkner}(2001)}]{bard_book}%
  \BibitemOpen
  \bibfield  {author} {\bibinfo {author} {\bibfnamefont {A.~J.}\ \bibnamefont
  {Bard}}\ and\ \bibinfo {author} {\bibfnamefont {L.~R.}\ \bibnamefont
  {Faulkner}},\ }\href@noop {} {\emph {\bibinfo {title} {Electrochemical
  Methods}}}\ (\bibinfo  {publisher} {J. Wiley \& Sons, Inc.},\ \bibinfo
  {address} {New York, NY},\ \bibinfo {year} {2001})\BibitemShut {NoStop}%
\bibitem [{\citenamefont {Barsoukov}\ and\ \citenamefont
  {Macdonald}(2005)}]{barsoukov2005}%
  \BibitemOpen
  \bibinfo {editor} {\bibfnamefont {E.}~\bibnamefont {Barsoukov}}\ and\
  \bibinfo {editor} {\bibfnamefont {J.~R.}\ \bibnamefont {Macdonald}},\ eds.,\
  \href@noop {} {\emph {\bibinfo {title} {Impedance spectroscopy: Theory,
  experiment and applications}}}\ (\bibinfo  {publisher} {J. Wiley \& Sons},\
  \bibinfo {year} {2005})\BibitemShut {NoStop}%
\bibitem [{\citenamefont {Song}\ and\ \citenamefont {Bazant}(2013)}]{song2013}%
  \BibitemOpen
  \bibfield  {author} {\bibinfo {author} {\bibfnamefont {J.}~\bibnamefont
  {Song}}\ and\ \bibinfo {author} {\bibfnamefont {M.~Z.}\ \bibnamefont
  {Bazant}},\ }\href@noop {} {\bibfield  {journal} {\bibinfo  {journal} {J.
  Electrochem. Soc.}\ }\textbf {\bibinfo {volume} {160}},\ \bibinfo {pages}
  {A15} (\bibinfo {year} {2013})}\BibitemShut {NoStop}%
\bibitem [{\citenamefont {Thomas-Alyea}(2008)}]{thomas2008}%
  \BibitemOpen
  \bibfield  {author} {\bibinfo {author} {\bibfnamefont {K.~E.}\ \bibnamefont
  {Thomas-Alyea}},\ }\href@noop {} {\bibfield  {journal} {\bibinfo  {journal}
  {ECS Transactions}\ }\textbf {\bibinfo {volume} {16}},\ \bibinfo {pages}
  {155} (\bibinfo {year} {2008})}\BibitemShut {NoStop}%
\bibitem [{\citenamefont {He\ss}\ and\ \citenamefont
  {Nov\'ak}(2013)}]{hess2013}%
  \BibitemOpen
  \bibfield  {author} {\bibinfo {author} {\bibfnamefont {M.}~\bibnamefont
  {He\ss}}\ and\ \bibinfo {author} {\bibfnamefont {P.}~\bibnamefont
  {Nov\'ak}},\ }\href@noop {} {\bibfield  {journal} {\bibinfo  {journal}
  {Electrochimica Acta}\ }\textbf {\bibinfo {volume} {106}},\ \bibinfo {pages}
  {149} (\bibinfo {year} {2013})}\BibitemShut {NoStop}%
\bibitem [{\citenamefont {Dresselhaus}\ and\ \citenamefont
  {Dresselhaus}(1981)}]{dresselhaus1981}%
  \BibitemOpen
  \bibfield  {author} {\bibinfo {author} {\bibfnamefont {M.}~\bibnamefont
  {Dresselhaus}}\ and\ \bibinfo {author} {\bibfnamefont {G.}~\bibnamefont
  {Dresselhaus}},\ }\href@noop {} {\bibfield  {journal} {\bibinfo  {journal}
  {Advances in Physics}\ }\textbf {\bibinfo {volume} {30}},\ \bibinfo {pages}
  {139} (\bibinfo {year} {1981})},\ \bibinfo {note} {also reprinted as Advances
  in Physics 51 (2002) 1}\BibitemShut {NoStop}%
\bibitem [{\citenamefont {Krishnan}\ \emph {et~al.}(2013)\citenamefont
  {Krishnan}, \citenamefont {Brenet}, \citenamefont {Machado-Charry},
  \citenamefont {Caliste}, \citenamefont {Genovese}, \citenamefont {Deutsch},
  ,\ and\ \citenamefont {Pochet}}]{krishnan2013}%
  \BibitemOpen
  \bibfield  {author} {\bibinfo {author} {\bibfnamefont {S.}~\bibnamefont
  {Krishnan}}, \bibinfo {author} {\bibfnamefont {G.}~\bibnamefont {Brenet}},
  \bibinfo {author} {\bibfnamefont {E.}~\bibnamefont {Machado-Charry}},
  \bibinfo {author} {\bibfnamefont {D.}~\bibnamefont {Caliste}}, \bibinfo
  {author} {\bibfnamefont {L.}~\bibnamefont {Genovese}}, \bibinfo {author}
  {\bibfnamefont {T.}~\bibnamefont {Deutsch}}, , \ and\ \bibinfo {author}
  {\bibfnamefont {P.}~\bibnamefont {Pochet}},\ }\href {\doibase
  10.1063/1.4850877} {\bibfield  {journal} {\bibinfo  {journal} {Applied
  Physics Letters}\ }\textbf {\bibinfo {volume} {103}},\ \bibinfo {pages}
  {251904} (\bibinfo {year} {2013})}\BibitemShut {NoStop}%
\bibitem [{\citenamefont {Bazant}(2011)}]{10.626}%
  \BibitemOpen
  \bibfield  {author} {\bibinfo {author} {\bibfnamefont {M.~Z.}\ \bibnamefont
  {Bazant}},\ }\href@noop {} {\emph {\bibinfo {title} {10.626 Electrochemical
  Energy Systems}}}\ (\bibinfo  {publisher} {Massachusetts Institute of
  Technology: MIT OpenCourseWare, http://ocw.mit.edu, License: Creative Commons
  BY-NC-SA},\ \bibinfo {year} {2011})\ \bibinfo {note} {lecture 11}\BibitemShut
  {NoStop}%
\bibitem [{\citenamefont {Dahn}(1991)}]{dahn1991}%
  \BibitemOpen
  \bibfield  {author} {\bibinfo {author} {\bibfnamefont {J.}~\bibnamefont
  {Dahn}},\ }\href@noop {} {\bibfield  {journal} {\bibinfo  {journal} {Phys.
  Rev. B}\ }\textbf {\bibinfo {volume} {44}},\ \bibinfo {pages} {9170}
  (\bibinfo {year} {1991})}\BibitemShut {NoStop}%
\bibitem [{\citenamefont {Nailmova}\ \emph {et~al.}(1995)\citenamefont
  {Nailmova}, \citenamefont {Guerart}, \citenamefont {Lelaurain},\ and\
  \citenamefont {Fateev}}]{nailmova1995}%
  \BibitemOpen
  \bibfield  {author} {\bibinfo {author} {\bibfnamefont {V.}~\bibnamefont
  {Nailmova}}, \bibinfo {author} {\bibfnamefont {D.}~\bibnamefont {Guerart}},
  \bibinfo {author} {\bibfnamefont {M.}~\bibnamefont {Lelaurain}}, \ and\
  \bibinfo {author} {\bibfnamefont {O.}~\bibnamefont {Fateev}},\ }\href@noop {}
  {\bibfield  {journal} {\bibinfo  {journal} {Carbon}\ }\textbf {\bibinfo
  {volume} {33}},\ \bibinfo {pages} {177} (\bibinfo {year} {1995})}\BibitemShut
  {NoStop}%
\bibitem [{\citenamefont {Zheng}\ \emph {et~al.}(1995)\citenamefont {Zheng},
  \citenamefont {Reimers},\ and\ \citenamefont {Dahn}}]{zheng1995}%
  \BibitemOpen
  \bibfield  {author} {\bibinfo {author} {\bibfnamefont {T.}~\bibnamefont
  {Zheng}}, \bibinfo {author} {\bibfnamefont {J.}~\bibnamefont {Reimers}}, \
  and\ \bibinfo {author} {\bibfnamefont {J.}~\bibnamefont {Dahn}},\ }\href@noop
  {} {\bibfield  {journal} {\bibinfo  {journal} {Phys. Rev. B}\ }\textbf
  {\bibinfo {volume} {51}},\ \bibinfo {pages} {734} (\bibinfo {year}
  {1995})}\BibitemShut {NoStop}%
\bibitem [{\citenamefont {Leger}\ \emph {et~al.}(1999)\citenamefont {Leger},
  \citenamefont {Argoul},\ and\ \citenamefont {Bazant}}]{leger1999}%
  \BibitemOpen
  \bibfield  {author} {\bibinfo {author} {\bibfnamefont {C.}~\bibnamefont
  {Leger}}, \bibinfo {author} {\bibfnamefont {F.}~\bibnamefont {Argoul}}, \
  and\ \bibinfo {author} {\bibfnamefont {M.~Z.}\ \bibnamefont {Bazant}},\
  }\href@noop {} {\bibfield  {journal} {\bibinfo  {journal} {J. Phys. Chem. B}\
  }\textbf {\bibinfo {volume} {103}},\ \bibinfo {pages} {5841} (\bibinfo {year}
  {1999})}\BibitemShut {NoStop}%
\bibitem [{\citenamefont {Bazant}\ and\ \citenamefont
  {Stone}(2000)}]{bazant2000}%
  \BibitemOpen
  \bibfield  {author} {\bibinfo {author} {\bibfnamefont {M.~Z.}\ \bibnamefont
  {Bazant}}\ and\ \bibinfo {author} {\bibfnamefont {H.~A.}\ \bibnamefont
  {Stone}},\ }\href@noop {} {\bibfield  {journal} {\bibinfo  {journal} {Physica
  D}\ }\textbf {\bibinfo {volume} {147}},\ \bibinfo {pages} {95} (\bibinfo
  {year} {2000})}\BibitemShut {NoStop}%
\bibitem [{\citenamefont {Stewart}\ and\ \citenamefont
  {Newman}(2008)}]{stewart2008}%
  \BibitemOpen
  \bibfield  {author} {\bibinfo {author} {\bibfnamefont {S.~G.}\ \bibnamefont
  {Stewart}}\ and\ \bibinfo {author} {\bibfnamefont {J.}~\bibnamefont
  {Newman}},\ }\href@noop {} {\bibfield  {journal} {\bibinfo  {journal}
  {Journal of the Electrochemical Society}\ }\textbf {\bibinfo {volume}
  {155}},\ \bibinfo {pages} {F13} (\bibinfo {year} {2008})}\BibitemShut
  {NoStop}%
\bibitem [{\citenamefont {Saravanan}\ \emph {et~al.}(2010)\citenamefont
  {Saravanan}, \citenamefont {Balaya}, \citenamefont {Reddy}, \citenamefont
  {Chowdari},\ and\ \citenamefont {Vittal}}]{saravanan2010}%
  \BibitemOpen
  \bibfield  {author} {\bibinfo {author} {\bibfnamefont {K.}~\bibnamefont
  {Saravanan}}, \bibinfo {author} {\bibfnamefont {P.}~\bibnamefont {Balaya}},
  \bibinfo {author} {\bibfnamefont {M.~V.}\ \bibnamefont {Reddy}}, \bibinfo
  {author} {\bibfnamefont {B.~V.~R.}\ \bibnamefont {Chowdari}}, \ and\ \bibinfo
  {author} {\bibfnamefont {J.~J.}\ \bibnamefont {Vittal}},\ }\href@noop {}
  {\bibfield  {journal} {\bibinfo  {journal} {Energy Environ. Sci.}\ }\textbf
  {\bibinfo {volume} {3}},\ \bibinfo {pages} {457} (\bibinfo {year}
  {2010})}\BibitemShut {NoStop}%
\end{thebibliography}%

\small

\section*{ Appendix A:  Iron Phosphate Simulations }

The general simulation framework of MPET is described in a companion paper~\cite{ferguson2012} and is beginning to be used by other researchers~\cite{dargaville2013porous}. In this Appendix, we give some of the parameters and simulation details not covered in the main text.

Cogswell demonstrated that single particle voltage profiles are always tilted due to coherency strain and that the overshoot from the standard potential on discharge varies with particle size.  This overshoot, which can be thought of as the half gap (i.e. the overpotential required to drive lithiation), depends on particle size since the surface wetting represents a larger percentage of the total volume for smaller particles.

As noted in the main text, the particles are modeled as effectively homogeneous (``pseudo capacitor approximation"~\cite{ferguson2012}), but with an equilibrium voltage profile that approximates coherency phase separation within  the particle~\cite{cogswell2012}.  We simply adjust the regular solution parameter $\Omega$ with particle size to place the unstable spinodal points (extrema of the chemical potential) at the size-dependent nucleation voltage predicted by Cogswell and Bazant~\cite{cogswell2013} (Eq. \ref{eq:nuc} below), as shown in Fig. \ref{figureM1}.  The value of $\Omega$ can be determined from an exact expression for the voltage gap between spinodal points in the regular solution model, 
\begin{equation}
\Delta V_{gap} = \frac{2k_BT}{e}\left[\sqrt{\tilde{\Omega}^2-2\tilde{\Omega}} -2\tanh^{-1}\left(\sqrt{1-\frac{2}{\tilde{\Omega}}}\right)\right].
\label{voltagegapvsOmega}
\end{equation}
In this approach, we neglect asymmetry between  charge and discharge in the single-particle properties.

\begin{figure}[htp]
\centering	
	\includegraphics[width=2.7in,keepaspectratio]{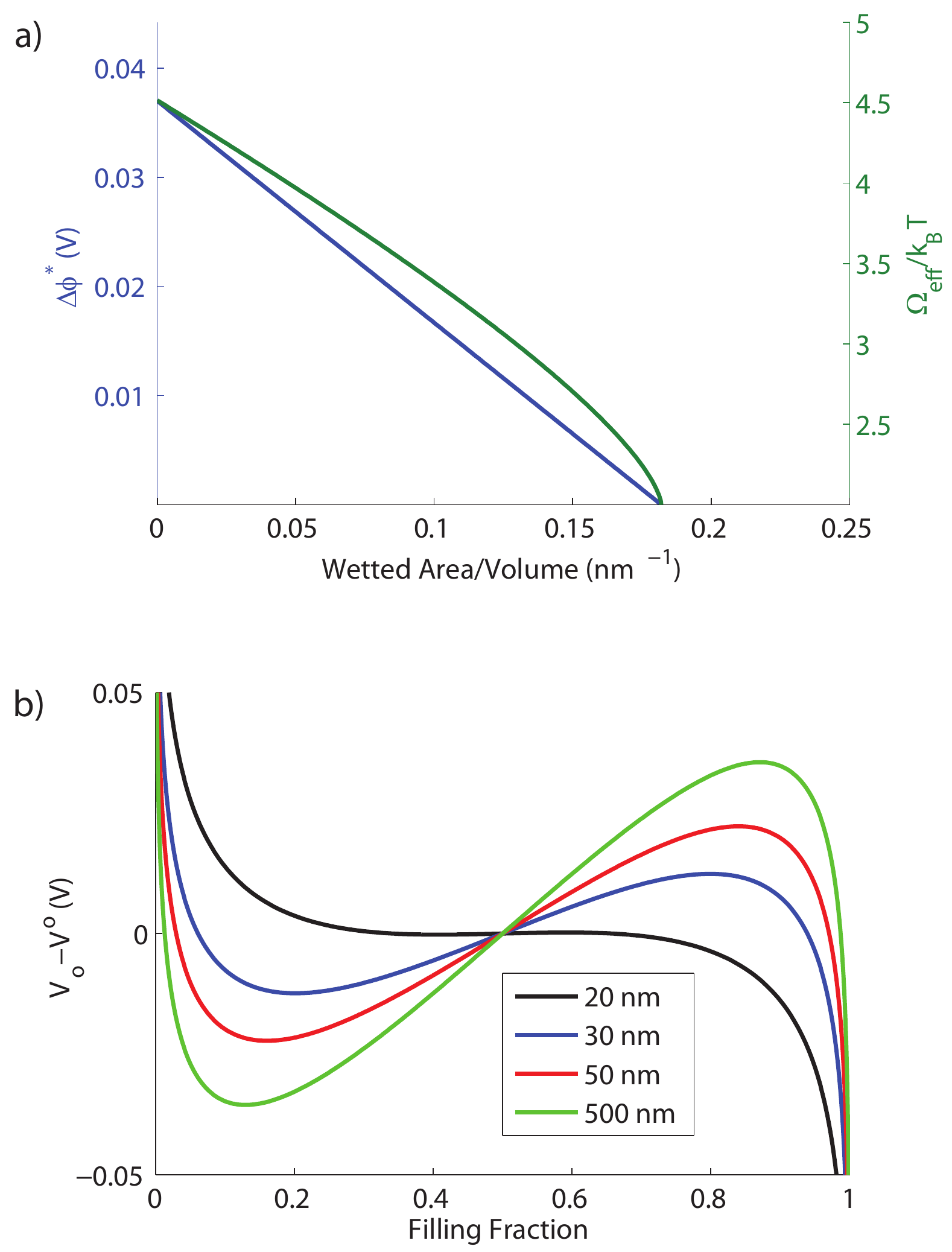}
\caption{ Size-dependent regular solution model to approximate coherent nucleation in nanoparticles. (a) Theoretical prediction of the critical nucleation voltage in LFP (blue), decreasing linearly with area to volume ratio~\cite{cogswell2013}, and the effective size-dependent regular solution parameter $\Omega_{eff}$.  (b) Equilibrium voltage profiles for different particle sizes from the size-dependent regular solution model fitted in (a).    
}
\label{figureM1}
\end{figure}

For purposes of calculating size-dependent nucleation barriers~\cite{cogswell2013}, the size of each representative particle in a finite volume of the porous electrode is sampled from a log-normal distribution.  The particle shapes are all assumed to be the C3 shape ~\cite{smith2012,cogswell2013}.  The particles are plate-like and all assumed to be 20 nm thick in the [010] direction ~\cite{smith2012,saravanan2010} . The wetted surface area to volume ratio is calculated via $A/V = 3.6338/L$, where $L$ is the size of the particle in the [100] direction. ~\cite{cogswell2013} .

Given the very slow charge/discharge rates in the experiments~\cite{dreyer2010}, the electrolyte properties are not very important since the electrolyte does not deplete.  For completeness, however, suitable numbers are chosen for transport properties.  An ambipolar diffusivity of 1.5x10$^{10}$m$^2$/s is used for the 1M LiPF$_6$ in EC/DEC electrolyte. ~\cite{stewart2008}.  A transference number does not seem to be available for this electrolyte, so a value of 0.35 is assumed, consistent with other typical battery electrolytes. ~\cite{handbookbatterymaterials} The electrode is assumed to be 50 $\mu$m long with a 25 $\mu$m separator.  The volume fraction of the active material is assumed to be 0.5, and the porosity is 0.4.

\begin{figure}[htp]
\centering	
	\includegraphics[width=2.7in,keepaspectratio]{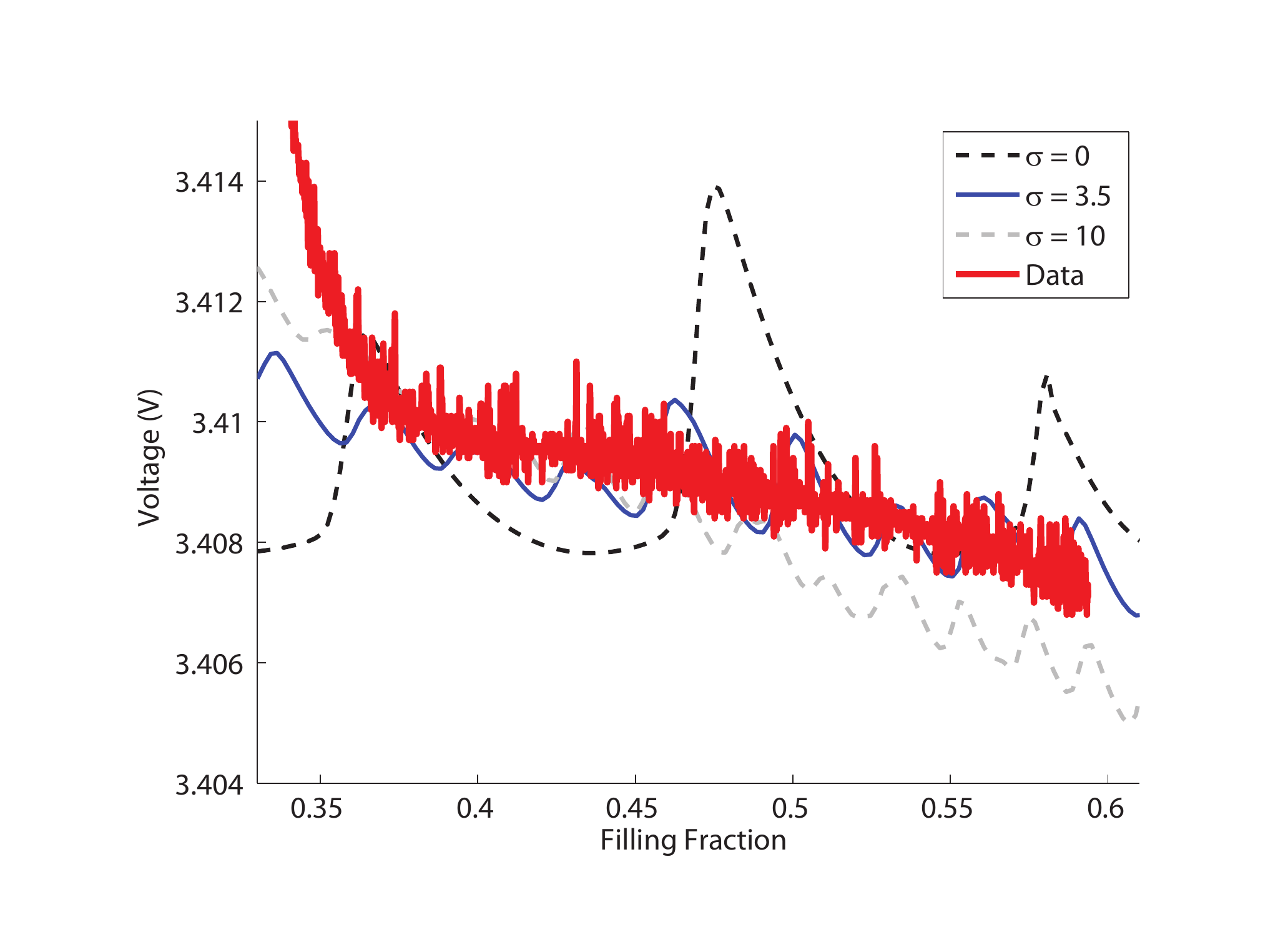}
\caption{  The raw data for the simulations of porous LFP low-rate discharge with different particle size distributions without any smoothing of the data.  In order to convey the effect of particle size and extract the voltage gaps  between charge and discharge, the same data is smoothed in Fig. 2(b).
}
\label{fig:nosmooth}
\end{figure}

The simulation is run by starting with a fully charged electrode, which is then discharged to a filling fraction of 0.2 at a C/10 rate.  The electrode is then relaxed by simulating a zero current for a long period of time (roughly 3.5 hours). The electrode is then discharged to a filling fraction of 0.7, relaxed, then charged back to 0.2.  When calculating the voltage gaps, the simulation results are smoothed to make the value consistent, to account for averaging over a large number of particles of different sizes that cannot be captured in our simulations with a relatively small set of discrete particle sizes.  The simulation results from Fig. 2(b) without smoothing are shown in Fig. ~\ref{fig:nosmooth}.  A constant contact resistance of 3.9 $\Omega\cdot$g is inferred by fitting the data, as one of only three fitting parameters, along with the mean and variance of the particle size in a log-normal distribution.

\section*{ Appendix B: Graphite Simulations }

The graphite free energy model of Bazant~\cite{10.626} described in the main text is adjusted so that a slow C/1000 MPET simulation (without transport limitation) fits open circuit voltage staircase (Fig.\ref{figure2}(a)), yielding the parameters $\Omega_a=3.4$k$_B$T, $\Omega_b=1.4$k$_B$T, and $V^o= 0.1366$ V.  The fourth parameter, $\Omega_c=30$k$_B$T,  assigns an extra energy to the homogeneous state that serves to push the particle toward stage II at intermediate filling fractions, but its precise value is relatively unimportant here.  Each representative layer is modeled as having its own reaction rate, determined by generalized Butler-Volmer kinetics~\cite{bazant2013}.  The slow discharge simulation (which allows the transport effects to be neglected) is used to fit the model parameters.  

Once the free energy parameters are obtained, a constant potential discharge simulation at $V=2 $mV is run for a full cell simulation with a lithium metal anode and a graphite cathode to simulate the experiments  of Harris et al.~\cite{harris2010}.  Transport effects on the lithium metal electrode are assumed to be negligible, and the porosity on that side is treated as unity to model the free electrolyte channel above the unrolled current collector.  The important transport effects are on the graphite side, where there is sharp electrolyte depletion from the separator to the intercalation front.  It is necessary to simulate a full electrode in order to obtain a realistic diffusivity from fitting. In a half cell neglecting the anode, large salt concentration gradients  form across the separator and lead to artificially small fitted diffusivities.

The separator thickness from images provided in the original paper is estimated to be 1.23 mm. ~\cite{harris2010}  The electrodes are assumed to be on the order of 1.2 cm (or 10 times the separator thickness).  The total length is not important since the coloration dynamics are observed in the first couple millimeters.  The diffusive time across the separator is on the order of one hour, which is rate limiting compared to the much faster reactions.   The exchange current densities  for both graphite and lithium metal are arbitarily set the inverse diffusion time across the separator, which corresponds to approximately 1.4 A/m$^2$ for 5 $\mu$m spherical particles. (Particle sizes of 5-20 $\mu$m are observed).  
The volume fraction of graphite in the electrode is assumed to be 0.8, and the Bruggeman relation is used to model porous transport effects~\cite{ferguson2012}.

\begin{figure*}[htp]
\centering	
	\includegraphics[width=5.5in,keepaspectratio]{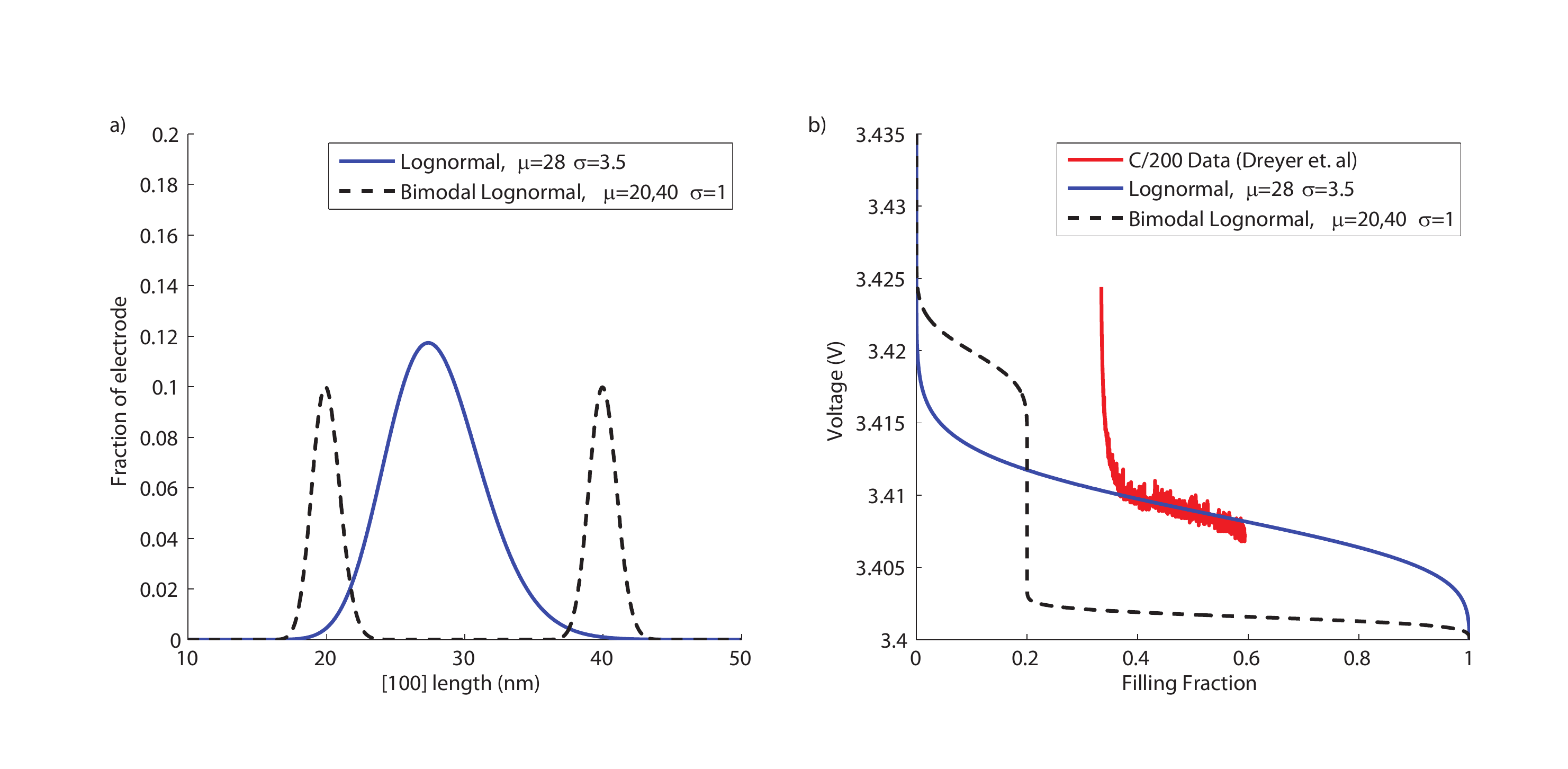}
\caption{ Theoretical dependence of the open circuit voltage during battery discharge on the particle size distribution of a phase separating porous electrode, in two cases with the same mean particle size. The single lognormal distribution  and a bimodal lognormal distribution shown on the left lead to the voltage profiles on the right.   The lognormal case is fitted to the experimental data of Dreyer et al.~\cite{dreyer2010} in Fig. 2  (red) with a minor adjustment of the standard potential. 
}
\label{fig:tilt}
\end{figure*}

\section*{ Appendix C:  Theory of the Tilted Voltage Plateau } 

Our MPET model reveals an unexpected effect of statistical variations in the active particle size on the open circuit voltage of a battery, namely a slight tilt of the voltage plateau between two stable phases.  This feature is present in published data for LFP at very low rates~\cite{dreyer2010}, but has apparently gone undetected until this work (Fig. \ref{figure1}). In this section, we provide a simple analytical theory of the voltage profile versus mean filling, $V(x)$, near open circuit conditions, based on the particle size distribution and the recent theory of size-dependent nucleation~\cite{cogswell2013}.

Let $\Delta \phi=V^\Theta-V-I R_s$ be the voltage drop relative to the formal reference potential of two-phase coexistence  $V^\Theta$, accounting for a small Ohmic loss from the overall series resistance $R_s$. For low rates, close to open circuit conditions, we neglect any concentration polarization within the electrode. Let $n(L)$ be the probability density function for the linear size $L$ of an active particle.   

Following Cogswell and Bazant~\cite{cogswell2013}, we assume that nucleation occurs on certain surfaces wetted by the new phase, e.g. the high density phase of LFP during battery discharge. The critical voltage for nucleation, $\Delta\phi^*$, which corresponds to the coherent miscibility limit, decreases linearly with the wetted area to volume ratio, $L/\beta$, 
\begin{equation}
\Delta \phi^* = \Delta \phi^*_\infty \left( 1 - \frac{L_c}{\beta L} \right)   \label{eq:nuc}
\end{equation}
where $\beta$ is a constant for a given particle shape (the volume to wetted area ratio per linear size, assumed to be the same for all particles), $\Delta \phi^*\infty$ is the critical voltage in the large-size limit (set by elastic strain energy), and $L_c$ is the critical particle size below which the nucleation barrier vanishes (due to the dominance of surface energy).  Simple analytical formulae are available to predict $\Delta \phi^*_\infty$ and $L_c$ from the fundamental thermodynamic properties of the active material (elastic constants, misfit strain, surface energies,  interfacial tension, miscibility limit, etc.), and in the case of LFP, the predicted values $\Delta \phi^*_\infty=37$ mV and $L_c=22$ nm lead to a remarkable data collapse of reported nucleation voltages for different mean particle sizes.  For the C3 shape of LFP particles~\cite{smith2012,cogswell2013}, the geometrical parameter is $\beta=(3.6338)^{-1}=0.2752$, if $L$ is the particle length in the [100] direction.

At a given voltage drop in the two-phase region, $0 < V < \Delta\phi^*_\infty$, near open circuit conditions, all particles with $L<L^*(\Delta\phi)$ are filled, and the others empty, where 
\begin{equation}
L^*(\Delta\phi) = \frac{\beta L_c}{1-\Delta\phi/\Delta\phi^*_\infty}
\end{equation}
is the size of the particles at the nucleation voltage, $L=L^*$, undergoing filling transformations. 
The mean filling fraction of the electrode is then given by
\begin{equation}
x = \frac{ \int_0^{L^*(\Delta\phi)} n(L) L^3 dL }{ \int_0^\infty n(L) L^3 dL }   \label{eq:volt}
\end{equation}
which is an implicit formula for the ``tilted voltage plateau" versus state of charge, $V(x)$. 
The voltage profile is generally nonlinear, but   its slope at a given point (inverse of the pseudo-capacitance) is given by
\begin{equation}
\frac{d\Delta\phi}{dx} = - \frac{v_p \left( 1-\Delta\phi/\Delta\phi^*_\infty \right)^5  \Delta\phi^*_\infty }
{ n(L^*(\Delta\phi)) \left( \beta L_c \right)^4 },
\end{equation}
where the third moment of the particle size distribution, $v_p = \int_0^\infty n(L) L^3 dL$, is an effective mean particle volume.  

Figure~\ref{fig:tilt} shows the effect of the shape of the particle size distribution on the profile of the open circuit voltage  versus state of charge. A unimodal distribution, such as a Gaussian or log-normal, leads to a tilted voltage plateau with a slightly nonlinear profile, as shown in Fig. 2, but a bimodal distribution leads to a voltage staircase with two tilted plateaus close to the mean voltages  where the two different particle sizes transform (at the size-dependent coherent miscibility limit).  Interestingly, this statistical mechanism for a voltage staircase is distinct from that of the graphite discussed in the main text, which has to do with transitions between three or more stable phases.  In principle, a very precise measurement of the tilted voltage plateau can be used to infer the particle size distribution, by solving an inverse problem given by Eq. (\ref{eq:volt}).

%This formula takes the following simple form for a log-normal distribution...
% -----
%TO DO:
%
%1) calculate for lognormal, and try to put theory curves on Fig 2b.
%
%2) make a new figure M2 plotting the tilted voltage plateau for lognormal and arbitrary bimodal and trimodal distributions (which will lead to a staircase like graphite!). Plot size distributions in (a) and voltage profiles in (b).

\end{document}